\documentclass[showpacs,floatfix,superscriptaddress]{revtex4}
\usepackage{graphicx}
\usepackage{epsfig}
\usepackage{bm}
\usepackage[utf8]{inputenc}
\usepackage{amssymb}
\usepackage{float}
\usepackage{amsmath}
\usepackage{multirow}
\usepackage{dcolumn}
\usepackage{latexsym}
\usepackage{subfig}
\usepackage{amsthm}
\usepackage{framed}
\usepackage{cancel}
\usepackage{mathtools}
\newcommand\myeq{\mathrel{\stackrel{\makebox[0pt]{\mbox{\normalfont\tiny $\tilde{R}\rightarrow\infty$}}}{=}}}
\newcommand\myeqq{\mathrel{\stackrel{\makebox[0pt]{\mbox{\normalfont\tiny $\tilde{R}\rightarrow\infty$}}}{\rightarrow}}}
\usepackage[colorlinks]{hyperref}
\usepackage[usenames,dvipsnames]{color}
\hypersetup{
     breaklinks=true,
    pdfstartview={FitH},    
    colorlinks=true,       
    linkcolor=blue,          
    citecolor=red,        
    filecolor=magenta,      
    urlcolor=blue,           
    anchorcolor=green,      
    linktocpage=true
}

\def\doi{http://doi.org}

\begin{document}

\title{ Higher-Dimensional MOG dark compact object: shadow behaviour in the light of EHT observations}

\author{Kourosh Nozari}
\email[]{knozari@umz.ac.ir}
\affiliation{Department of Theoretical Physics, Faculty of Sciences, University of Mazandaran,\\
47416-95447, Babolsar, Iran}

\author{Sara Saghafi}
\email[]{s.saghafi@umz.ac.ir}
\affiliation{Department of Theoretical Physics, Faculty of Sciences, University of Mazandaran,\\
47416-95447, Babolsar, Iran}

\author{Ali Mohammadpour}
\email[]{alimohammadpour9898@gmail.com}
\affiliation{Department of Theoretical Physics, Faculty of Sciences, University of Mazandaran,\\
47416-95447, Babolsar, Iran}

\begin{abstract}
Consideration of extra spatial dimensions is motivated by the unification of gravity with other interactions, the achievement of the ultimate framework of quantum gravity, and fundamental problems in particle physics and cosmology. Much attention has been focused on the effect of these extra dimensions on the modified theories of gravity. Analytically examining astrophysical phenomena like black hole shadows is one approach to understand how extra dimensions would affect the modified gravitational theories. The purpose of this study is to derive a higher-dimensional metric for a dark compact object in STVG theory and then examine the behaviour of the shadow shapes for this solution in STVG theory in higher dimensions. We apply the Carter method to formulate the geodesic equations and the Hamilton-Jacobi method to find photon orbits around this higher-dimensional MOG dark compact object. We investigate the effects of extra dimensions and the STVG parameter $\alpha$ on the black hole's shadow size. Next, we compare the shadow radius of this higher-dimensional MOG dark compact object to the shadow size of the supermassive black hole M87*, which has been realized by the Event Horizon Telescope (EHT) collaborations, in order to restrict these parameters. We find that extra dimensions in the STVG theory typically lead to a reduction in the shadow size of the higher-dimensional MOG dark compact object, whereas the effect of parameter $\alpha$ on this black hole's shadow is suppressible.  Remarkably, given the constraints from EHT observations, we find that  the shadow size of the four-dimensional MOG dark compact object lies in the confidence levels of the EHT data. Finally, we investigate the
issue of acceleration bounds in higher-dimensional MOG dark compact object in confrontation with EHT data of  M87*.
\end{abstract}

\pacs{04.50.Kd, 04.70.-s, 04.20.Ha, 04.25.dg, 04.50.-h, 04.50.Gh, 97.60.Lf}

\maketitle

\enlargethispage{\baselineskip}
\tableofcontents

\section{Introduction}
Dark compact objects, from a phenomenological perspective, represent a wide range of astronomical objects, such as white dwarfs, neutron stars, and black holes. From a theoretical perspective, it is possible to predict dark compact objects within the framework of extended gravity theories and in scenarios involving particle physics beyond the standard model \cite{Cardoso:2019rvt}. Gravitational waves detected by LIGO/Virgo observations recently demonstrated the occurrence of binary black hole mergers, see \cite{LIGOScientific:2017bnn} and references therein. In addition, the Event Horizon Telescope (EHT) demonstrated the presence of supermassive black holes in the Milky Way and M87 galactic centers \cite{EventHorizonTelescope:2021srq,EventHorizonTelescope:2022xqj} and references therein. Thus, it is not surprising that new species of compact objects will be discovered by future developments in the fields of very long baseline interferometry (VLBI) \cite{APhaseSensitiveInterferometer:1985} and gravitational wave astronomy. However, the study of dark compact objects from a mathematical perspective is interesting, because it can assist in comprehending how and at what limits a dark compact object tends to be a black hole by increasing its compactness. As mentioned, the Event Horizon Telescope (EHT) collaboration's pioneering achievement of constructing the first pictures of the shadows of the supermassive black holes M87* (see \cite{EventHorizonTelescope:2021srq}  and references
therein)  and Sgr A* (see \cite{EventHorizonTelescope:2022xqj}  and references
therein) has opened a new chapter in the physics of black holes and dark compact objects and enhanced our understanding of these mysterious celestial objects. Since a black hole does not emit light, its event horizon, that is, the surface at which matter or radiation cannot pass through, is not directly observable. Instead, what we can see is the ''shadow" of a black hole, a dark region surrounded by light that results in the gravitational lensing phenomenon around the event horizon~ \cite{Perlick:2004tq,Cosmpactness:2022,Relativisticimage:2009,ImagesDistortionHypothesis:2024,Cosmologicalconstant:2022,Ishihara:2016,
Tsukamoto:2017,Ishihara:2017,Virbhadra:2000,Claudel:2001,Virbhadra:2002,Virbhadra:1998,Tsukamotoa:2023,Kumaran:2023}. Much attempts have been made to measure properly in order to obtain high-resolution images since the release of the shadow images of M87* and Sgr A*~ \cite{Goddi:2016qax}. To achieve the needed resolution, theoretical efforts to study physics of black hole, in particular, shadow behavior in different theories of gravitation, has become of increasing significance. Novel models of theoretical shadow for upcoming observations are provided in this case by numerical and analytical studies and analyses of the geometric shape of different black hole spacetimes. More than spacetime characteristics ~\cite{Johannsen:2010ru,Cunha:2017eoe} and the observer's place, the parameters of black hole; angular momentum, mass , and electric charge determine the shadow's shape and size \cite{deVries:1999tiy}. The shadow shape is a circle and a perfect one for black holes which are non-rotating. But black holes which are rotating may have got non-trivial shadow patterns due to the angular momentum parameter ~\cite{Chandrasekhar:1984siy}.

Even though Albert Einstein's General Theory of Relativity (GR) has made significant progress in explaining observations and predicting astounding events, it is still unable to fully explain gravitational interaction and the events that correspond with it in the universe. Some instances of the shortcomings of GR are the reproduction of the rotation curves of nearby galaxies \cite{Sofue:2000jx,Sofue2016}, mass profiles of galaxy clusters \cite{Ettori2013,Voigt2006}, inherent singularities at the center of black holes, etc. Furthermore, the positively accelerated expansion of the Universe at late time is explained by GR only if the cosmological constant factor $\Lambda$ is present \cite{Weinberg1989,Peebles2003}. Restructuring the geometric part of General Relativity (GR) using various ways is an intriguing reform strategy. One such approach is the so-called MOdified Gravity (MOG), a Scalar-Tensor-Vector (STVG) theory developed and proposed by John W. Moffat to describe gravitational interaction \cite{Moffat2006}. The gravitational effects of spacetime in the MOG setup are expressed by a massive vector field $\phi_\mu$ in addition to three scalar fields: the mass of the vector field $\tilde{\mu}$, the effective gravitational constant $G$, and the vector field coupling $\xi$. In addition to being compatible with cosmological observational data \cite{Davari2021} and Planck 2018 data ~\cite{Moffat2021a}, the MOG theory has achieved various successes in explaining astrophysical phenomena. These include explaining the dynamics of galaxy clusters without dark matter and the rotation curves of many galaxies ~\cite{Brownstein2006a,Brownstein2006b,Brownstein2007,Moffat2013,Moffat2014,Moffat2015a}. Furthermore, the framework of MOG theory is currently experiencing the release of a number of black hole solutions, including non-rotating and rotating ones ~\cite{Moffat2015b}, even with extra dimensions \cite{Cai2021}, cosmological solutions \cite{Roshan2015,Jamali2018,Davari2021}, and non-stationary solutions for inhomogeneity distributions of mass–energy in spacetime ~\cite{Perez2019}. Additionally, a great deal of theoretical and observational work has been done to comprehend the characteristics of the MOG theory and how it functions in various contexts ~\cite{Moffat2021b,Guo2018,Moffat2020,Mureika2016,Saghafi2021,Saghafi:2021wzx,Perez:2017spz,Shukla:2022sti,Hussain:2015cga,Sharif:2017owq,Lee:2017fbq,Kolos:2020ykz,
Hu:2022lek,John:2019was,Camilloni:2024,rosa:2023}. Ref. ~\cite{Moffat:2018jmi} offers an interesting exploration of the solution characterizing the regular rotating and non-rotating MOG dark compact object. Ref. ~\cite{Sau:2022afl} examines the shadow behavior of the regular rotating and non-rotating MOG dark compact object.

The notion of higher dimensions was initially put forward by Nordstrom in 1914 ~\cite{Nordstrom:1914ejq}. His theory states that Using four-dimensional spacetime as a surface in a five-dimensional spacetime, one may unify the gravitational and electromagnetic fields. Currently, the primary motivations of the vast amount of research on the extra dimensions are the quantization of gravitational interaction, the unification of gauge and gravitational force among elementary particles, the the cosmological constant problem, and Higgs mass hierarchy problem. This is why the Kaluza-Klein (KK) theory \cite{Kaluza:1921tu,Klein:1926tv} unifies non-Abelian gauge fields or  electromagnetic  that feature weak and strong forces with gravitational interaction by introducing a compact space built from compact extra dimensions with a specific compactification scale. Additionally, as a famous theory,a candidate for quantum gravity, the string theory (M-theory) has ten or eleven compact dimensions \cite{Witten:1995ex,Schwarz:1995jq}. There are some thoughts for large additional dimensions of the order of millimeter in addition to these compact extra dimensions with the extension up to the order of the Planck length. The Arkani-Hamed-Dimopoulos-Dvali (ADD) braneworld model \cite{Arkani-Hamed:1998jmv,Arkani-Hamed:1998sfv} has provided a novel path for the study of extra dimensions by using large extra dimensions to solve the Higgs mass hierarchy problem. It is important to remember that the striking aspect of these large extra dimensions is that future accelerator, astrophysical, and even tabletop studies will be able to measure their effects. It is surprising to learn that string theory can incorporate the ADD model ~\cite{Antoniadis:1998ig}. The Randall-Sundrum (RS) braneworld model \cite{Randall:1999ee,Randall:1999vf} proposes the use of warped extra dimensions in addition to compact and large higher dimensions to solve the Higgs mass hierarchy problem. Apart from the aforementioned extra dimensions, there exist theories with unlimited volume extra dimensions such as the Dvali-Gabadadze-Porrati (DGP) braneworld scenario ~\cite{Dvali:2000hr}, wherein the spacetime is not four-dimensional even at very low energies and the higher dimensions are not wrapped and not compact. These theories, which alter gravity at large distances because the existence of infinite volume additional dimensions, are possibilities for solving the cosmological constant problem \cite{Dvali:2000xg,Dvali:2002pe}. Refs. \cite{Gabadadze:2003ii,Shifman:2009df,Perez-Lorenzana:2005fzz,Rubakov:2001kp} provide some thorough evaluations of higher-dimensional models. A variety of techniques and strategies have been used in the context of physics of black holes to expand and study alternative solutions for black holes in arbitrary dimensions \cite{Emparan:2008eg,Kanti:2004nr}. One such technique is the Tangherlini method \cite{Tangherlini:1963bw}, which generalizes the solution for Schwarzschild to $n$ dimensions.

 Physicists concentrate on detecting extra dimensions in high energy regimes. Prospective instruments for investigating the impact of regimes of strong gravity related to black holes with extra dimensions and extra dimensions are the Large Hadron Collider at CERN and future colliders \cite{Allanach:2002gn,Agashe:2006hk,Franceschini:2011wr,Deutschmann:2017bth,Strominger:1996sh,Harris:2004xt,Antoniadis:2000vd}. Furthermore, evidence for the existence of extra dimensions comes from spectroscopic tests \cite{Zhou:2014xbw,Luo:2006ck,Luo:2006ad}, the existence of hydrogen atoms in extra dimensions \cite{Burgbacher:1999sha,Caruso:2012daf,Shaqqor:2009cha}, and theories to solve the proton radius puzzle \cite{Wang:2013fma,Dahia:2015bza,Zhi-gang:2007swh}. However, the detection of gravitational waves (GW) by the LIGO/Virgo collaborations \cite{LIGOScientific:2016aoc} and the aforementioned images obtained by the EHT collaborations are two recent successes in the search for black hole strong field regimes. Some signs of additional dimensions are in the GW detection, with specific details regarding the associated amplitude and fluctuation mode dynamics. As a result, a lot of research has been done to uncover this physics \cite{Cardoso:2016rao,Yu:2016tar,Visinelli:2017bny,Kwon:2019gsa}; for a thorough analysis, refer to Ref. \cite{Yu:2019jlb}. EHT has now opened up novel opportunities for the search for extra dimensions. The authors of recent seminal publications \cite{Vagnozzi:2019apd,Belhaj:2020mlv,Tang:2022hsu} discovered significant limitations on warped and compact extra dimensions beyond the M-theory and RS model, derived from EHT observations. Thus, as we aim to do in this study, the EHT data can be used to investigate whole kinds of higher dimensions in general and notice if they can be observed. In this sense, it appears that extra dimensions have an impact on black hole shadows by decreasing the size of the shadow in a variety of black hole models and gravitational theories \cite{Amarilla:2011fx,Eiroa:2017uuq,Papnoi:2014aaa,Singh:2017vfr,Amir:2017slq,Belhaj:2020rdb}. Nevertheless, there is still much to learn about the precise impact of higher dimensions on black hole shadows, which is the subject of ongoing research.

Furthermore, due to additional degrees of freedom that comes from modified theories of gravity, black hole shadow size and shape may vary. Thus, determining the size and shape of black hole shadows could help evaluate black hole metrics parameters and test alternative theories of gravity. The question of black hole shadow has been the subject of a great deal of research to determine whether degrees of freedom affect the behavior of the shadow \cite{Perlick:2021aok}. Here are a few instances: Einstein-Maxwell-dilaton gravity black holes and their shadows \cite{Amarilla:2013sj,vag:2023,vag1:2023,vag2:2023,vag3:2023,vag4:2023,vag5:2023,Wei:2013kza}, in Chern-Simons modified gravity \cite{Amarilla:2010zq}, the Kerr-Newman family of solutions of the Einstein-Maxwell equations and its shadow behavior is studied in Refs. \cite{Bardeen:1973tla,Takahashi:2004xh}; the black hole with NUT-charges and its shadow \cite{Chakraborty:2013kza,Grenzebach:2014fha}; the apparent shape of the Sen black hole ~\cite{Hioki:2008zw,Dastan:2016bfy,Younsi:2016azx}; the non-commutative geometry inspired and quantum-corrected black holes and their shadows ~\cite{Wei:2015dua,Sharif:2016znpx,Saghafi:2022pmey}; Einstein-Born-Infeld black holes and its shadow~ \cite{Atamurotov:2015xfa}; Ayon-Beato-Garcia black hole and also, rotating Hayward and rotating Bardeen regular black holes and their shadows~ \cite{Abdujabbarov:2016hnw} and hairy black holes \cite{Cunha:2015yba,Cunha:2016bjh}; colliding and multi-black holes and shadows \cite{Nitta:2011nin,Yumoto:2012kz}; investigating shadows of rotating black holes in $f(R)$ gravity \cite{Dastan:2016vhb}, conformal Weyl gravity ~\cite{Mureika:2016efo}, and Einstein-dilaton-Gauss-Bonnet black holes ~\cite{Cunha:2016wzk}; studying Shadow of wormholes and naked singularities \cite{Nedkova:2013msa,Ohgami:2015nra,Ortiz:2015rma}, as well as chaotic shadow of a non-Kerr rotating compact object with quadrupole mass moment and a magnetic dipole ~\cite{Wang:2018eui,Wang:2017qhh}, and black holes with exotic matter~ \cite{Tinchev:2015apf,Abdujabbarov:2015pqp,Singh:2017xle,Huang:2016qnl}.

In this article, we aim to develop a higher-dimensional metric for a dark compact object in STVG theory, derive its black hole solutions, and investigate its horizons in different dimensions. Afterward, we explore the optical properties of dark compact object, such as deflection, shadow size, photon sphere, and effective potential ($V_{eff}$). Finally by calculating the deflection angle and shadow size in different dimensions, and comparing them with observational data obtained from M87* and Sgr A*, we evaluate whether our universe is an extra dimensional universe. Additionally, we will assess the level of acceptance of STVG theory as an alternative gravitational theory.

This paper is based as follows. In section \ref{sec2} first the Modified Gravity theory is introduced and then we introduce the line element of the higher-dimensional dark compact object in this theory. In section \ref{sec3} we study the shadow behavior, energy emission rate, and deflection angle of the STVG dark compact object in MOG theory with higher dimensions, and the impact of parameter $\alpha$ and dimensions on the shadow and deflection angle of the dark compact object is analyzed. In section \ref{sec4} we constrain the parameter $\alpha$ by comparing the shadow size of the suppermassive black hole M87* with the shadow size of the STVG dark compact object. In section
\ref{sec5} we examin the issue of acceleration bounds in higher-dimensional MOG dark compact object's spacetime background. Finally, Section \ref{sec6} would discuss summary, conclusion and our main results.

\section{MOG and its higher-dimensional regular static spherically symmetric dark compact object}\label{sec2}

The STVG theory's overall action takes the form of \cite{Moffat2006}:
\begin{equation}\label{tact}
S=S_{GR}+S_{M}+S_{\phi}+S_{S}\,,
\end{equation}
where $S_{GR}$ is the Einstein-Hilbert action, $S_{M}$ is the action of all possible matter sources, $S_{\phi}$ is the action of the (spin $1$ graviton) vector field $\phi^{\mu}$ possessing the mass $\tilde{\mu}$ as one of the scalar fields in the theory, and $S_{S}$ is the action of three scalar fields, which can be expressed as follows
\begin{equation}\label{sgr}
S_{GR}=\frac{1}{16\pi}\int d^{D}x\sqrt{-g}\frac{1}{G}R\,,
\end{equation}
\begin{equation}\label{sphi}
S_{\phi}=-\int d^{D}x\sqrt{-g}\left(\frac{1}{4}B^{\mu\nu}B_{\mu\nu}+V_{1}(\phi)\right)\xi\,,
\end{equation}
\begin{equation}\label{ss}
\begin{split}
S_{S} & =\int d^{D}x\sqrt{-g}\left[\frac{1}{G^{3}}\left(\frac{1}{2}g^{\mu\nu}\nabla_{\mu}G\nabla_{\nu}G-V_{2}(G)\right)
+\frac{1}{\tilde{\mu}^{2}G}\left(\frac{1}{2}g^{\mu\nu}\nabla_{\mu}\tilde{\mu}\nabla_{\nu}\tilde{\mu}-V_{3}(\tilde{\mu})\right)\right.\\
& \left.+\frac{1}{G}\left(\frac{1}{2}g^{\mu\nu}\nabla_{\mu}\xi\nabla_{\nu}\xi-V_{4}(\xi)\right)\right]\,,
\end{split}
\end{equation}
in which $D$ represents the dimension of spacetime, $g_{\mu\nu}$ is the background metric tensor and $g$ is the corresponding determinant, $R$ is the Ricci scalar constructed by contracting $R_{\mu\nu}$ as the Ricci tensor, $G$ is a scalar field in the setup, which known as the enhanced Newtonian parameter, $\xi$ is third scalar field in the setup as the vector field coupling, $V_{1}(\phi)$, $V_{2}(G)$, $V_{3}(\tilde{\mu})$, and $V_{4}(\xi)$ are the corresponding potentials of the vector field $\phi^{\mu}$, and three scalar fields $G$, $\tilde{\mu}$, and $\xi$, respectively, and $B_{\mu\nu}=\partial_{\mu}\phi_{\nu}-\partial_{\nu}\phi_{\mu}$, and also $\nabla_{\mu}$ stands for the covariant derivative in the spacetime.

In the STVG theory, $T_{\mu\nu}={}^{(M)}T_{\mu\nu}+{}^{(\phi)}T_{\mu\nu}+{}^{(S)}T_{\mu\nu}$ is the total stress-energy tensor, in which the stress-energy tensor of matter sources is ${}^{(M)}T_{\mu\nu}$, the stress-energy tensor of the scalar fields is ${}^{(S)}T_{\mu\nu}$, and the stress-energy tensor of the vector field is
\begin{equation}\label{setphi}
{}^{(\phi)}T_{\mu\nu}=-\frac{1}{4}\left(B_{\mu}^{\,\,\,\sigma}B_{\nu\sigma}-\frac{1}{4}g_{\mu\nu}B^{\sigma\lambda}B_{\sigma\lambda}\right)\,,
\end{equation}
for which $V_{1}(\phi)=0$. One can find the full field equations of the STVG framework by variation of the action $S$ concerning the inverse of the metric tensor, which yields \cite{Moffat2006}
\begin{equation}\label{ffe}
G_{\mu\nu}+G\left(\nabla^{\gamma}\nabla_{\gamma}\frac{1}{G}g_{\mu\nu}-\nabla_{\mu}\nabla_{\nu}\frac{1}{G}\right)=8\pi G T_{\mu\nu}\,,
\end{equation}
in which the Einstein tensor is defied as $G_{\mu\nu}=R_{\mu\nu}-\frac{1}{2}g_{\mu\nu}R$ and we have set $c=1$.

\subsection{Regular MOG static spherically symmetric dark compact object}

The line element of the regular MOG static spherically symmetric dark compact object were found under the following assumptions \cite{Moffat:2018jmi}
\begin{itemize}
  \item The vector field is massless, i.e., $\tilde{\mu}=0$, since one can prove that for MOG compact objects, e.g., black holes possessing horizons, the mass of the vector field in the setup is zero.
  \item The enhanced Newtonian parameter $G$ is defined as a constant depending on the free dimensionless parameter $\alpha$ so that $G=G_{N}(1+\alpha)$ where $G_{N}$ is the Newtonian constant. Furthermore, the gravitational source charge of the vector field is $Q_{g}=\sqrt{\alpha G_{N}}M$ where $M$ is the source mass. Here, we set $G_{N}=1$.
  \item The vector field coupling is set to unity, i.e., $\xi=1$.
  \item The matter-free field equations of STVG setup is considered since the MOG dark compact object is a vacuum solution of the framework.
\end{itemize}
The above assumptions result in $S_{M}=S_{S}=0$ and consequently, we have ${}^{(M)}T_{\mu\nu}={}^{(S)}T_{\mu\nu}=0$. Thus, the field equations \eqref{ffe} now reduce to the following form
\begin{equation}\label{rfe}
G_{\mu\nu}=8\pi(1+\alpha){}^{(\phi)}T_{\mu\nu}\,.
\end{equation}

To solve a higher-dimensional static spherically symmetric compact object in the STVG, the line element is typically assumed to have the following form:
\begin{equation}\label{le}
ds^{2}=-h(r)dt^{2}+\frac{1}{f(r)}dr^{2}-r^{2}d\Omega_{D-2}^{2}\,,
\end{equation}

where $d\Omega^{2}_{D-2}=d\theta_{1}^{2}+\sin^{2}\theta_{1}d\theta_{1}^{2}+...+\prod_{i=1}^{D-3}\sin^{2}\theta_{i}d\theta_{D-2}^{2}$ represents the line element on the $(D-2)$-dimensional unit sphere.
\begin{equation}\label{new-metric}
f(r)=h(r)=1-\frac{(D-2)mr^{2(D-3)}\omega_{D-2}}{8\pi(r^{2(D-3)}+\frac{(D-2)^{2}m^{2}\alpha(1+\alpha)\omega^{2}_{D-2}}{256G^{2}\pi^{2}})^{\frac{3}{2}}}
+\frac{(D-3)(D-2)Gq^{2}r^{2(D-3)}\omega^{2}_{D-2}}{32\pi^{2}(r^{2(D-3)}+\frac{(D-2)^{2}m^{2}\alpha(1+\alpha)\omega^{2}_{D-2}}{256G^{2}\pi^{2}})^{2}}
\end{equation}

where $m$ and $q$ are defined by
\begin{equation}
\label{14}
m\equiv \frac{16\pi GM}{(D-2)\omega_{_{D-2}}},\qquad q\equiv \frac{8\pi Q}{\sqrt{2(D-2)(D-3)}\omega_{_{D-2}}},
\end{equation}
and $\omega_{_{D-2}}=\frac{2\pi^{\frac{D-1}{2}}}{\Gamma (\frac{D-1}{2})}$ means the area of a $(D-2)$-dimensional unit sphere.

 It is obvious that the metric function Eq.~(\ref{new-metric}) turns back to the MOG dark compact object solution given by Moffat~\cite{Moffat:2018jmi} for the case of $D=4$. The MOG dark compact object possesses a critical value for $\alpha$ as $\alpha_{crit}=0.674$ \cite{Moffat:2018jmi}, so that for $\alpha\leq\alpha_{crit}$ it has two horizons. It is worth mentioning that the (spin $1$ graviton) vector field produces a repulsive gravitational force, which prevents the collapse of the MOG dark compact object to a MOG black hole with horizon.

Setting $\alpha=0$ in the line element \eqref{le} recovers the Schwarzschild-Tangherlini black hole in GR. Moreover, the asymptotic behavior of the higher-dimensional MOG compact object in the limit of  $r\rightarrow\infty$ is deduced as follows
\begin{equation}\label{fleabl}
f(r)\approx 1-\frac{2(1+\alpha)m}{r^{D-3}}+\frac{\alpha(1+\alpha)Gq^{2}}{r^{2(D-3)}}\,.
\end{equation}
When, $\alpha\leq\alpha_{crit}$, the two horizons of the regular higher-dimensional MOG static spherically symmetric dark compact object in the limit of $r\rightarrow\infty$ can be found as
\begin{equation}\label{hrz}
r_{\pm}=\Big(M+M\alpha \pm \sqrt{M^{2}+M^{2}\alpha}\Big)^{\frac{1}{D-3}}\,.
\end{equation}
Where $r_{-}$ is the inner horizon called the Cauchy horizon and $r_{+}$ is the outer horizon called the event horizon. In addition, when $\alpha =0$, the two horizons are merged into the event horizon of the Schwarzschild-Tangherlini black hole~\cite{Tangherlini:1963,polchinski:1998}. When, $\alpha>\alpha_{crit}$, there is a naked regular MOG static spherically symmetric dark compact object with no horizon. On the other hand, approaching the source, i.e., $r\rightarrow 0$, the 4D-MOG dark compact object behaves to the form
\begin{equation}\label{fleabs}
f(r)\approx 1-\frac{r^{2}}{M^{2}}\left(\frac{2\sqrt{1+\alpha}-\sqrt{\alpha}}{(1+\alpha)\alpha^{\frac{3}{2}}}\right)\,.
\end{equation}
Therefore, the spacetime metric of the MOG dark compact object is regular so that $f(0)=1$. Additionally, one can verify that the Kretschmann scalar $R^{\mu\nu\lambda\sigma}R_{\mu\nu\lambda\sigma}$ in addition to the Ricci scalar $R$ in the spacetime metric are regular at $r=0$.

For the static spherically symmetric system, the gravitational redshift $z$ at the asymptotic distance $r$ to an observer is gathered as follows
\begin{equation}\label{grs}
z(r)=\frac{1}{\sqrt{f(R)}}-1\,,
\end{equation}
where the MOG dark compact object's radius is $R$. The gravitational redshift of the compact object becomes infinite on the horizon $r_{+}$ for $\alpha<\alpha_{crit}$, and has a finite value for $\alpha>\alpha_{crit}$ in the limit of $r\rightarrow\infty$. However, one expects the regular MOG dark compact object in order to be sufficiently dark in order to be consistent with binary X-ray observations based on observational data., so that $\alpha\sim \alpha_{crit}$ \cite{Moffat:2018jmi}.

In spacetime dimensions $D=4, 5, 6, 7$, we plot the graph of the metric function $f(r)$ with respect to $r$ for various values of $\alpha$. It should be noted that the Schwarzschild-Tangherlini black hole case in Einstein's gravity is represented by the parameter $\alpha=0$. This image illustrates how, for $D=4,5, 6, 7$, the event horizon radius $r_{\rm H}$ decreases as the parameter $\alpha$ decreases. Furthermore, before we reach the extreme situation, we may observe that the higher-dimensional MOG dark compact object always has one more horizon than the Schwarzschild-Tangherlini black hole. Furthermore, by increasing the spacetime dimension both the event horizon and the inner Cauchy horizon radius decrease.

\begin{figure}[htb]
\centering
{\includegraphics[width=0.48\textwidth]{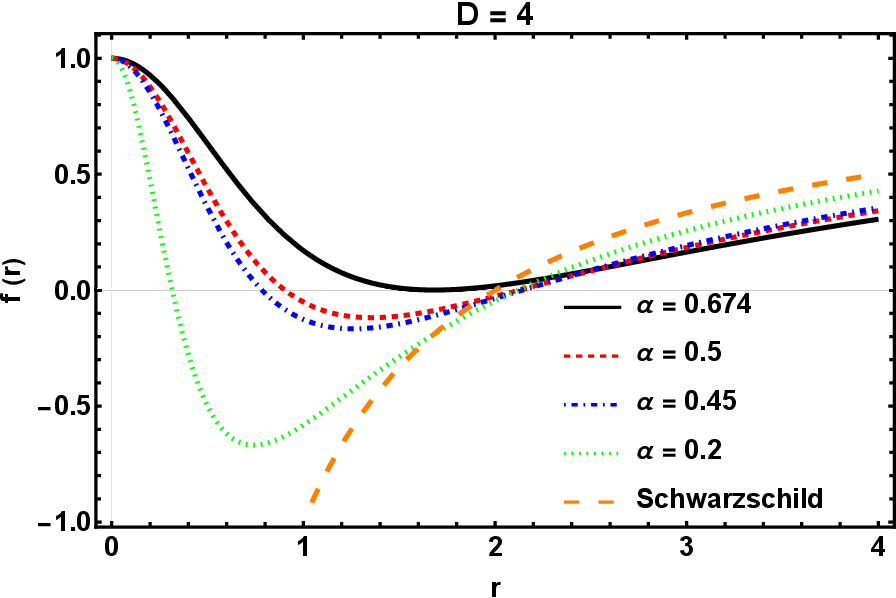}}
\,\,\,
{\includegraphics[width=0.499\textwidth]{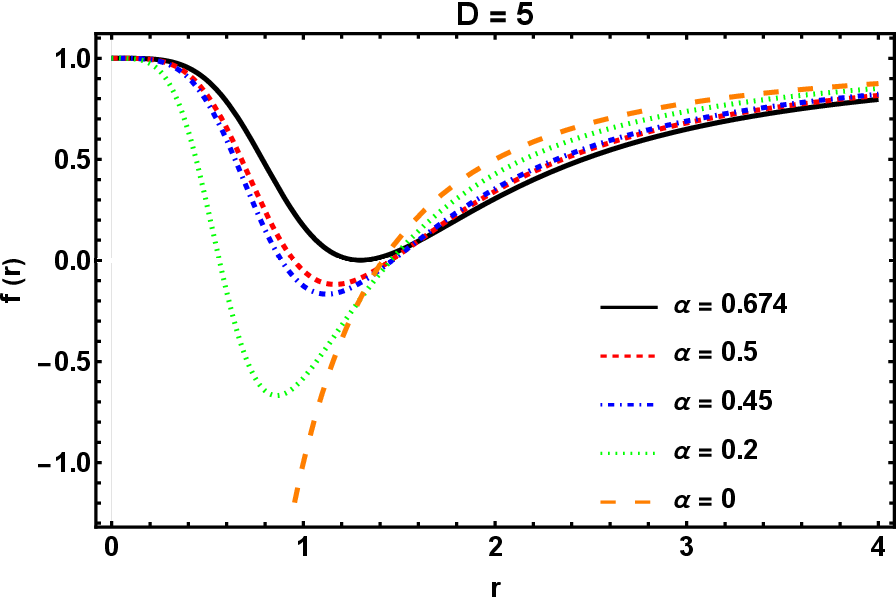}} \\
{\includegraphics[width=0.49\textwidth]{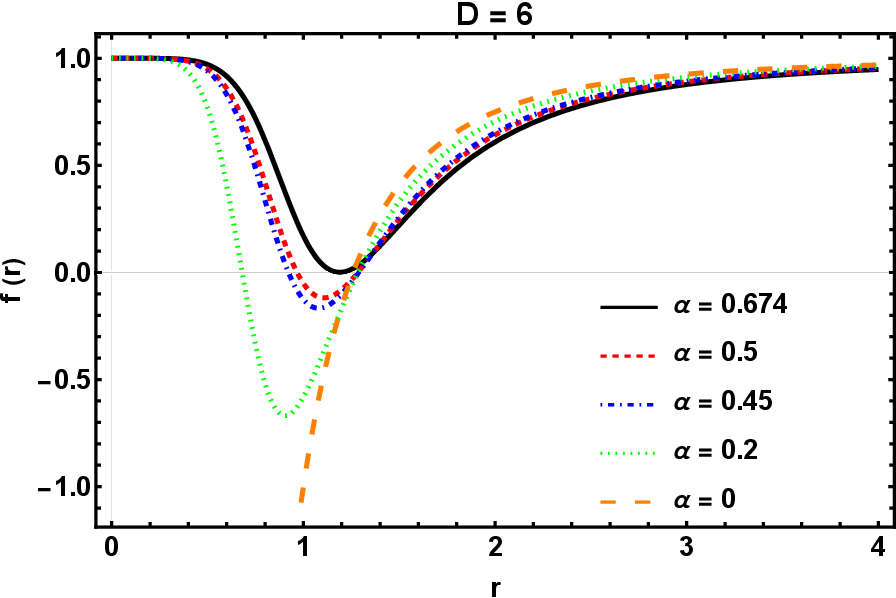}}
\,\,\,
{\includegraphics[width=0.49\textwidth]{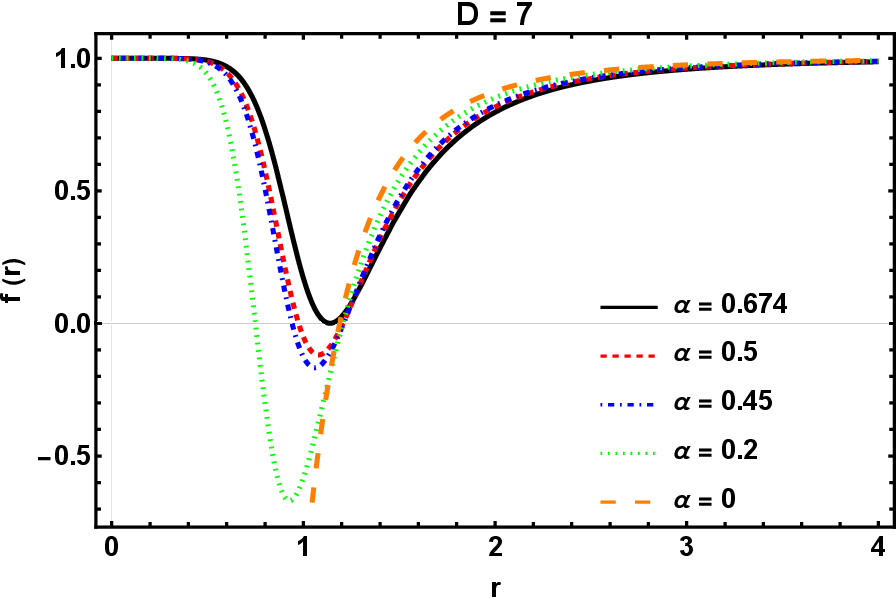}}
\caption{\label{figure4}\small{\emph{Function $f(r)$ with respect to $r$ for different values of $\alpha$, where the spacetime dimensions are taken to be $D=4,5, 6, 7$, respectively. Here we set $M=1$.}}}
\end{figure}

\section{Shadow behaviour of the higher-dimensional MOG dark compact object}\label{sec3}
The gravitational field of a black hole deflects some of the light that would otherwise reach the observer, when it passes in front of it. The black hole's apparent shape is the shadow's boundary, which is created when photons fall into the black hole and form a dark area known as the shadow. The general formulae needed to determine the deflection angle, energy emission rate, and shadow shape for the higher-dimensional general ansatz (\ref{le})—which requires examining a test particle's motion in spacetime—are provided in this section. Then we apply higher-dimensional MOG dark compact object line element on the formulas deduced in this section.

In MOG theory with extra dimensions, we analyze the shadow and deflection angle of the dark compact object by taking arbitrary  values for  the STVG parameter, $\alpha= 0, 0.2, 0.45, 0.5$, and $0.674$. In \cite{Sau:2022afl}, the authors consider regular MOG dark compact objects, using nonlinear field equations to determine the effects of $\phi_{\mu}$ for $\alpha > 0.674$ by adopting the ADM mass, here we examine the case for $\alpha\leq\alpha_{crit}$.

\subsection{Form of the Effective potential}

Initially, we aim to examine the effective potential behaviour of the dark compact object with arbitrary dimensions.
To do this, the Lagrangian of the test particle:
\begin{equation}\label{lagrangian}
\tilde{\mathcal{L}}=\frac{1}{2}g_{ab}\dot{x}^{a}\dot{x}^{b}\,,
\end{equation}
in which dot is the derivative with respect to affine parameter $\tau$. The canonically conjugate momentum components corresponding with the Eq. (\ref{le}) are as follows
\begin{equation}\label{pt}
P_{t}=h(r)\dot{t}=E\,,
\end{equation}
\begin{equation}\label{pr}
P_{r}=\frac{1}{f(r)}\dot{r}\,,
\end{equation}
\begin{equation}\label{pte}
P_{\theta_{i}}=r^{2}\sum_{i=1}^{D-3}\prod_{D=1}^{i-1}\sin^{2}(\theta_{D})\dot{\theta}_{i}\,,
\end{equation}
\begin{equation}\label{pph}
P_{\theta_{D-2}}=r^{2}\prod_{i=1}^{D-3}\sin^{2}(\theta_{i})\dot{\theta}_{D-2}=L\,,
\end{equation}
where $i=1,2,3,...,D-3$, and $L$ is the angular momentum and $E$ is the energy of the test particle. We apply the Carter approach to study the geodesic equations and the Hamilton-Jacobi method to analyze photon orbits around the black hole. We need to make these methods general to extra dimensions for our work. The Hamilton-Jacobi method in extra dimensions has the following form
\begin{equation}\label{HJ1}
\frac{\partial S}{\partial\tau}=-\frac{1}{2}g^{ab}\frac{\partial S}{\partial x^{a}}\frac{\partial S}{\partial x^{b}}\,,
\end{equation}
We have the following by putting the general arbitrary dimensions ansatz. (\ref{le}) into Eq. (\ref{HJ1}),
\begin{equation}\label{HJ2}
\begin{split}
-2 \frac{\partial S}{\partial\tau}=-\frac{1}{h(r)}\left(\frac{\partial S_t}{\partial t}\right)^{2}+f(r)\left(\frac{\partial S_r}{\partial r}\right)^{2}+\sum_{i=1}^{D-3}\frac{1}{\left(r^{2}\prod_{D=1}^{i-1}\sin^{2}\theta_{D}\right)}\left(\frac{\partial S_{\theta_{i}}}{\partial \theta_{i}}\right)^{2}+\frac{1}{\left(r^{2}\prod_{i=1}^{D-3}\sin^{2}\theta_{i}\right)}\left(\frac{\partial S_{\theta_{D-2}}}{\partial\theta_{D-2}}\right)^{2}\,.
\end{split}
\end{equation}
By considering a separable solution for Jacobi action, the action can be written as
\begin{equation}\label{jaction}
S=\frac{1}{2} m^{2}\tau-Et+L\theta_{D-2}+S_{r}(r)+\sum^{D-3}_{i=1}S_{\theta_{i}}(\theta_{i})\,,
\end{equation}
Since the photon is the test particle used to analyze the shadow behavior of black holes, $m$ is set to zero.
Also by inserting the action (\ref{jaction}) into Eq. (\ref{HJ2}) we derive the following equations
\begin{equation}\label{HJ3}
\begin{split}
0 & =\left\{\frac{E^{2}}{h(r)}-f(r)\left(\frac{\partial S_r}{\partial r}\right)^{2}-\frac{1}{r^{2}}\left(\frac{L^{2}}{\prod_{i=1}^{D-3}\sin^{2}\theta_{i}}+\mathcal{K}-\prod_{i=1}^{D-3}L^{2}\cot^{2}
\theta_{i}\right)\right\}\\
& -\left\{\frac{1}{r^{2}}\left(\sum_{i=1}^{D-3}\frac{1}{\prod_{D=1}^{i-1}\sin^{2}\theta_{D}}
\left(\frac{\partial S_{\theta_{i}}}{\partial \theta_{i}}\right)^{2}-\mathcal{K}+\prod_{i=1}^{D-3}L^{2}\cot^{2}
\theta_{i}\right)\right\}\,,
\end{split}
\end{equation}
where $\mathcal{K}$ is the Carter constant. The following set of equations can be obtained after some manipulations
\begin{equation}\label{HJ4}
r^{4}f^{2}(r)\left(\frac{\partial S_r}{\partial r}\right)^{2}=r^{4}\frac{f(r)}{h(r)}E^{2}-r^{2}\left(L^{2}+\mathcal{K}\right)f(r)\,,
\end{equation}
\begin{equation}\label{HJ5}
\sum_{i=1}^{D-3}\frac{1}{\prod_{D=1}^{i-1}\sin^{2}\theta_{D}}\left(\frac{\partial S_{\theta_{i}}}{\partial \theta_{i}}\right)^{2}=\mathcal{K}-\prod_{i=1}^{D-3}L^{2}\cot^{2}\theta_{i}\,.
\end{equation}
Finally, the equation of motion for photon (null geodesic within the higher-dimentional spacetime (\ref{le})) can be derived as follows, by utilizing Eqs. (\ref{HJ4}) and the canonically conjugate momentum components (\ref{pt})-(\ref{pph})
\begin{equation}\label{td}
\dot{t}=\frac{E}{f(r)}\,,
\end{equation}
\begin{equation}\label{rd}
r^{2}\dot{r}=\pm\sqrt{\mathcal{R}}\,,
\end{equation}
\begin{equation}\label{ted}
r^{2}\sum_{i=1}^{D-3}\prod_{D=1}^{i-1}\sin^{2}(\theta_{D})\dot{\theta}_{i}=\pm\sqrt{\Theta_{i}}\,,
\end{equation}
\begin{equation}\label{phd}
\dot{\theta}_{D-2}=\frac{L}{r^{2}\prod_{i=1}^{D-3}\sin^{2}\theta_{i}}\,,
\end{equation}
The ''$-$'' and ''$+$'' represent respectively the ingoing and outgoing radial directions of photon motion. 
Also in the spacetime , we have the photon's motion as follows which can be derived by Eqs. (\ref{td})-(\ref{phd}).
\begin{equation}\label{RT}
\mathcal{R}=r^{4}\frac{f(r)}{h(r)}E^{2}-r^{2}\left(L^{2}+\mathcal{K}\right)f(r)\,,\quad \Theta_{i}=\mathcal{K}-\prod_{i=1}^{D-3}L^{2}\cot^{2}\theta_{i}\,.
\end{equation}

The radial null geodesic equation is rewritten (\ref{rd}) and yields the effective potential,which has the following form:
\begin{equation}\label{potential1}
V_{eff}=\frac{f(r)}{r^{2}}(\mathcal{K}+L^{2})-\frac{f(r)}{h(r)}E^{2}
\end{equation}
where $\mathcal{K}$, $E$ and $L$ are the constants of motion.

The black hole's apparent shape is defined by its photons' unstable circular orbits. The photon sphere radius $r_{0}$ corresponds to the greatest value of the effective potential at a specific distance, as shown in the equations below.

\begin{equation}\label{twoeq}
V_{eff}\big|_{r_{0}}=\frac{dV_{eff}}{dr}\bigg|_{r_{0}}=0\,,\quad \mathcal{R}\big|_{r_{0}}=\frac{d\mathcal{R}}{dr}\bigg|_{r_{0}}=0\,.
\end{equation}

Therefore, The photon sphere radius $r_{0}$ corresponds to the black hole's maximal effective potential in spacetime (\ref{le}). With arbitrary dimensions,
\begin{equation}\label{r0}
r_{0}h'(r_{0})-2h(r_{0})=0\,,
\end{equation}
where a prime stands for radial derivative.

By inserting Eq. (\ref{new-metric}) into Eq. (\ref{potential1}) we can derive the effective potential of higher-dimensional MOG dark compact object as follows
\begin{equation}\label{potential2}
V_{eff}=1-\frac{(D-2)mr^{2(D-4)}\omega_{_{D-2}}}{8\pi(r^{2(D-3)}+\frac{(D-2)^{2}m^{2}\alpha(1+\alpha)\omega^{2}_{D-2}}{256G^{2}\pi^{2}})^{\frac{3}{2}}}+\frac{(D-3)(D-2)Gq^{2}r^{2(D-4)}\omega^{2}_{D-2}}{32\pi^{2}(r^{2(D-3)}+\frac{(D-2)^{2}m^{2}\alpha(1+\alpha)\omega^{2}_{D-2}}{256G^{2}\pi^{2}})^{2}}(\mathcal{K}+L^{2})-E^{2}
\end{equation}

\begin{figure}[htb]
\centering
\subfloat[\label{ep2} for $\alpha=0.5$]{\includegraphics[width=0.49\textwidth]{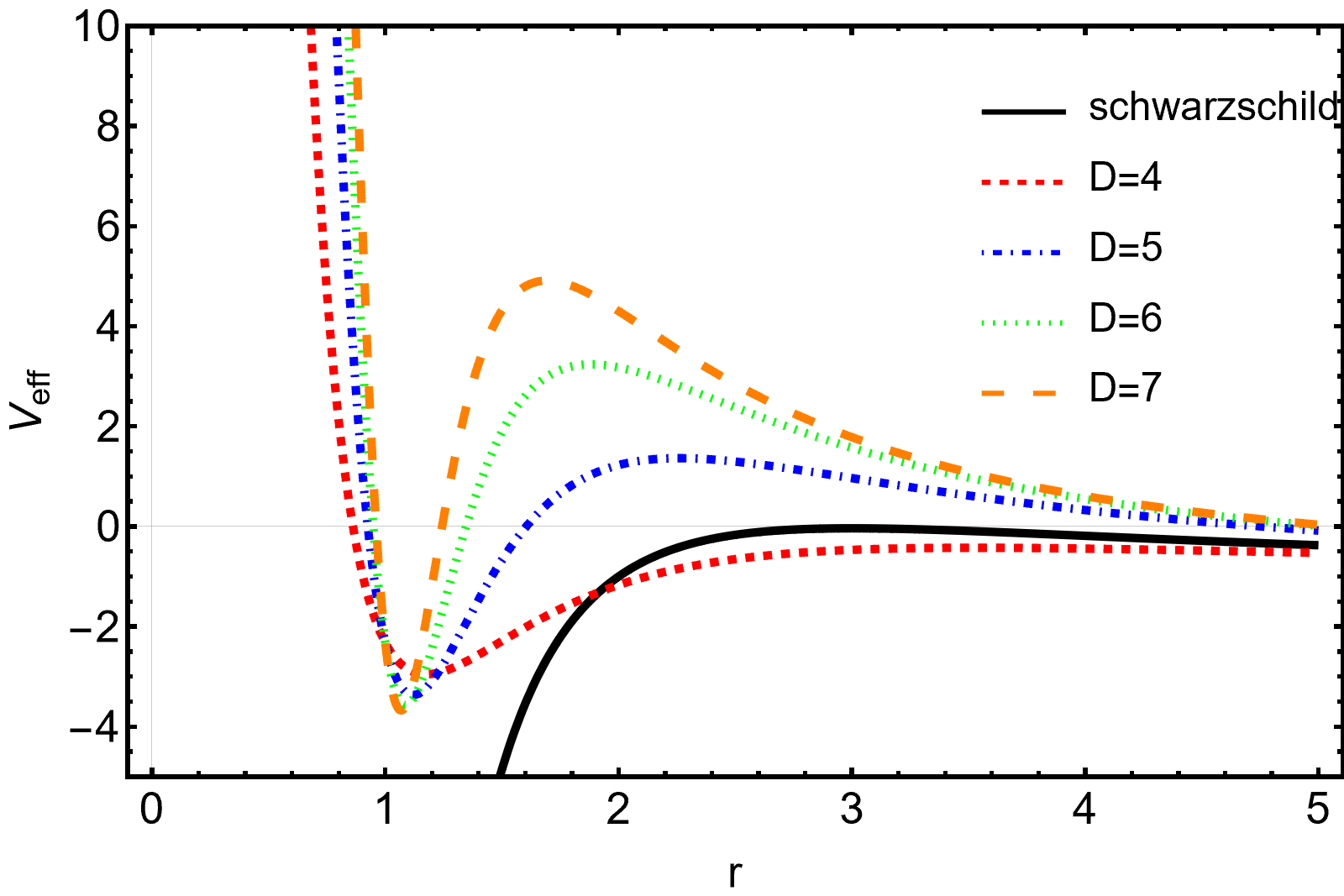}} \\
\subfloat[\label{ep3} for $D=4$]{\includegraphics[width=0.49\textwidth]{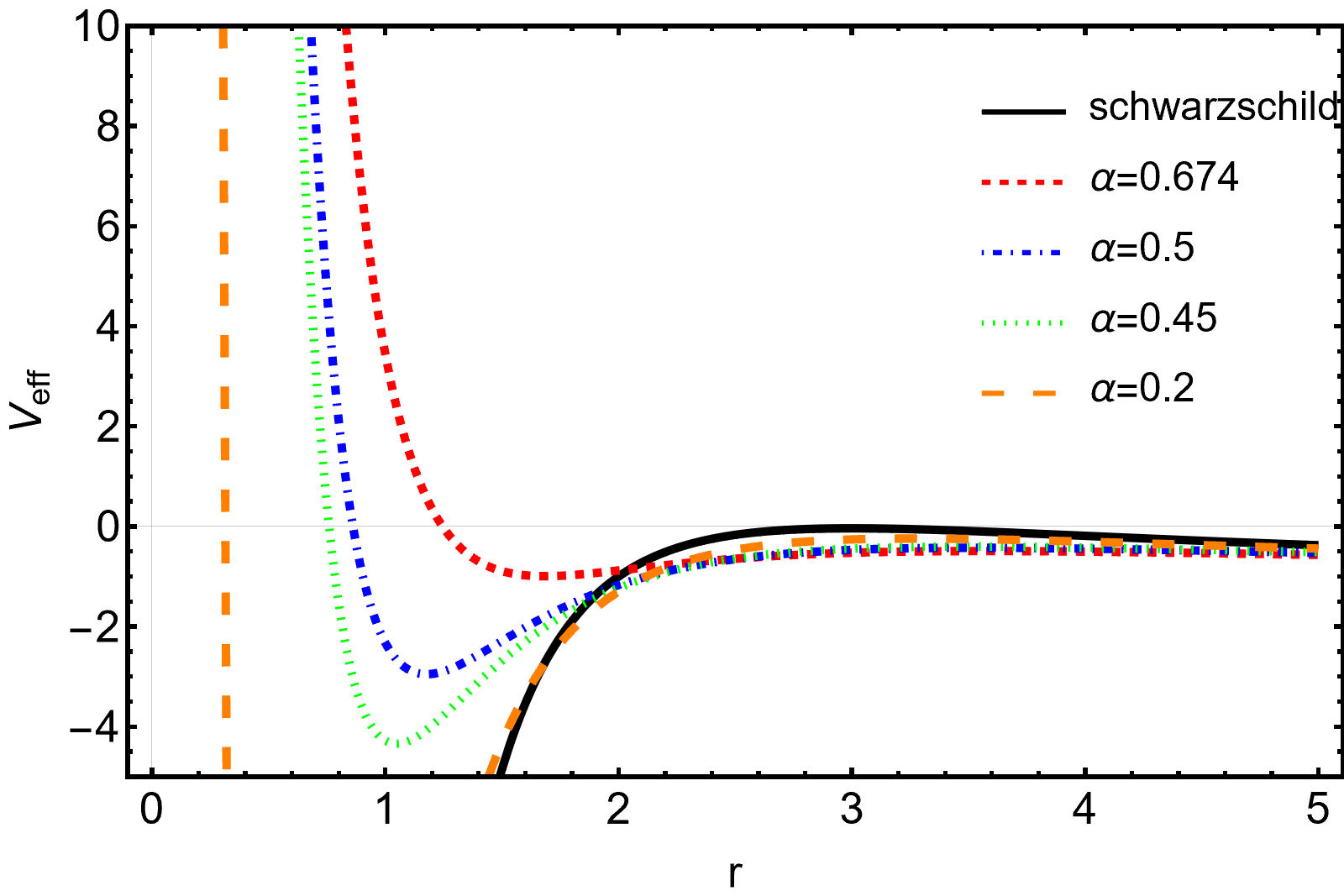}}
\,\,\,
\subfloat[\label{ep4} for $D=6$]{\includegraphics[width=0.49\textwidth]{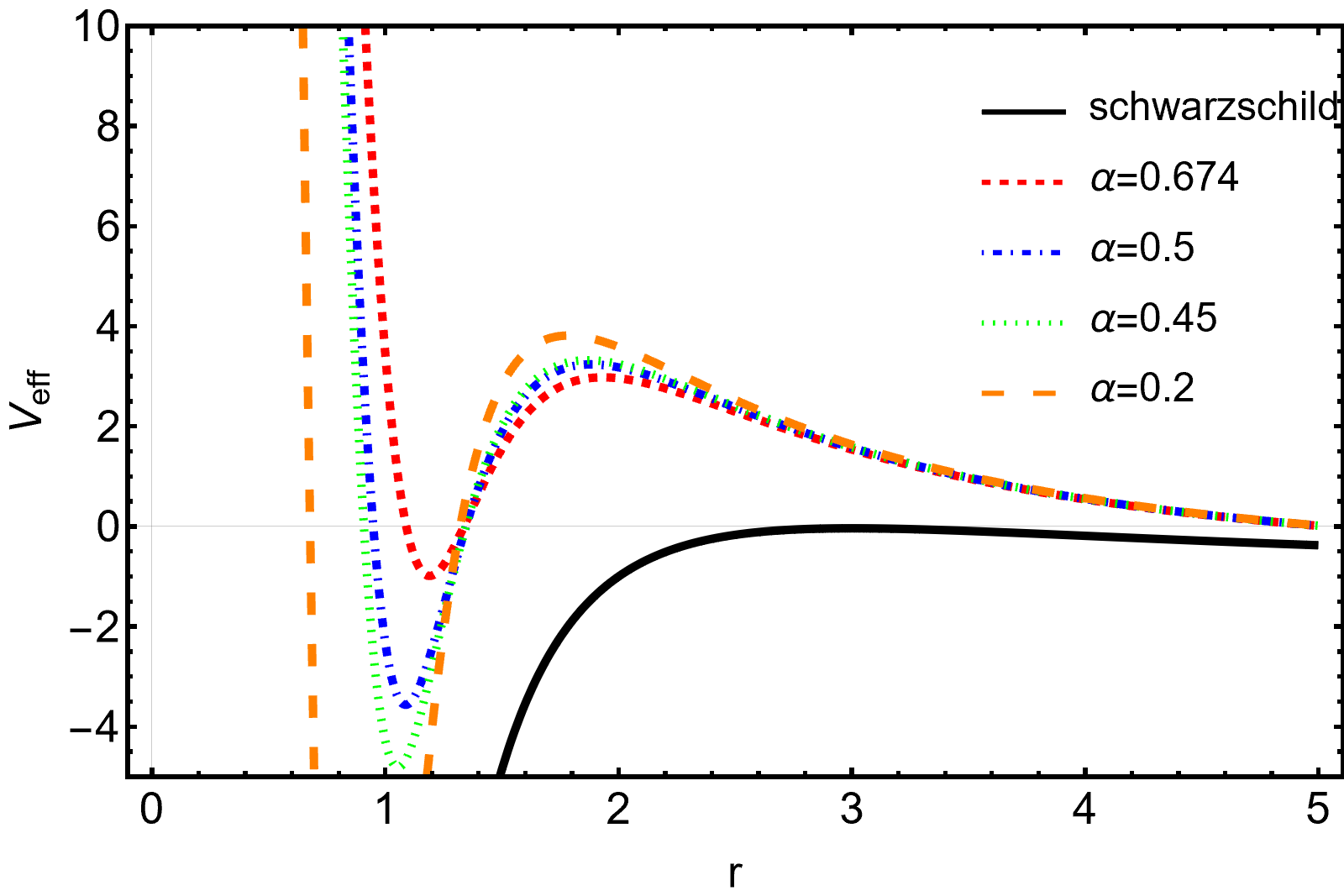}}
\caption{\label{Fig2}\small{\emph{The figure of the radial variation of effective potential of the dark compact object with extra dimensions in STVG theory for various values of $D$ and $\alpha$,where we set $L=5$ and $E=\mathcal{K}=M=1$.}}}
\end{figure}

For various values of $D$ and $\alpha$, the behavior of the effective potential for the higher dimensional MOG dark compact object is shown in Fig. \ref{Fig2} as a function of radial coordinate $r$. The photon sphere radius $r_0$, which corresponds to each value of $D$ and $\alpha$, is where the effective potential in this figure peaks. As we can observe from Fig. \ref{ep2}, the higher-dimensional MOG dark compact object's effective potential grows as $D$ increases for a constant value of $\alpha$. Furthermore, we discover from Figs. \ref{ep3} and \ref{ep4} that raising $\alpha$ for $D = 4,6$ causes the dark compact object's effective potential to be amplified. By examining these figures it is obvious that Higher dimensions have greater impact than $\alpha$. in the MOG dark compact object by increasing the spacetime dimensionality. Figure \ref{Fig2} illustrates how $D$ and $\alpha$ impact the shadow boundary of the MOG dark compact object by increasing the spacetime dimensionality. This is because the location of the black hole's maximum effective potential, describes the black hole's shadow boundary.

\subsection{Geometric shapes of shadow}\label{GEO}

We want to determine the shadow form and size of the black hole in the arbitrary-dimensional spacetime \eqref{le} in this section. First, we define two impact parameters, $\xi$ and $\eta$, in order to accomplish this end. The characteristics of photons close to black holes can be described by these impact parameters as functions of the constants of motion $E$, $L$, and $\mathcal{K}$. They provide the following definitions:

\begin{equation}\label{xieta1}
\xi=\frac{L}{E}\,,\qquad\eta=\frac{\mathcal{K}}{E^{2}}\,.
\end{equation}
The effective potential and function $\mathcal{R}$ can be rewritten in terms of impact parameters, as
\begin{equation}\label{veffR}
V_{eff}=E^{2}\left\{\frac{f(r)}{r^{2}}\left(\eta+\xi^{2}\right)-\frac{f(r)}{h(r)}\right\}\,,\qquad
\mathcal{R}=E^{2}\left\{r^{4}\frac{f(r)}{h(r)}-r^{2}f(r)\left(\eta+\xi^{2}\right)\right\}\,.
\end{equation}
Putting Eq. \eqref{veffR} into Eq. \eqref{twoeq} yields the following equation for two unknowns $\xi$ and $\eta$.
\begin{equation}\label{xieta2}
\eta+\xi^{2}=\frac{r_{0}^{2}}{2f(r_{0})+r_{0}f'(r_{0})}\left\{4\left(\frac{f(r_{0})}{h(r_{0})}\right)+r_{0}
\left(\frac{f'(r_{0})h(r_{0})-f(r_{0})h'(r_{0})}{h^{2}(r_{0})}\right)\right\}\,.
\end{equation}

Consequently, using Eq. \eqref{xieta2},The photon sphere radius from Eq. \eqref{r0} produces the quantity $\eta+\xi^{2}$. It is evident that the quantity $\eta+\xi^{2}$ has the dimension of the length square, while $r_{0}$ has the dimension of length.

Usually $\lambda$ and $\psi$  are used as celestial coordinates to describe the geometric shape of the shadow as perceived by the observer's frame \cite{Vazquez:2003zm}. These coordinates are

\begin{equation}\label{ab}
\lambda=\lim_{r_{o}\rightarrow\infty}\left(\frac{r_{o}^{2}P^{(\theta_{n-2})}}{P^{(t)}}\right)\,,\qquad
\psi=\lim_{r_{o}\rightarrow\infty}\left(\frac{r_{o}^{2}P^{(\theta_{i})}}{P^{(t)}}\right)\,,
\end{equation}
where $\left[P^{(t)},P^{(\theta_{D-2})},P^{(\theta_{i})}\right]$ are its vi-tetrad momentum elements and $r_{o}$ is the distance between the observer and the black hole. $\psi=\pm\sqrt{\eta}$ and $\lambda=-\xi$ are found on the equatorial plane. As a result, we can achieve the following result

\begin{equation}\label{Rs}
R_{s}^{2}\equiv\eta+\xi^{2}=\lambda^{2}+\psi^{2}\,,
\end{equation}

where the shadow radius in celestial coordinates is denoted by $R_{s}$. The geometric shape of the shadow for non-rotating (static) black holes is a circle of radius $R_{s}$.

We will now demonstrate the geometrical shape of the MOG dark compact object's shadow in higher dimensions on the observer's sky using the celestial coordinates that were presented above. To begin with, data for the black hole's related $ r_{eh}, r_{0}$, and $\sqrt{\eta+\xi^{2}}$ are gathered. For the higher-dimensional MOG dark compact object, the photon sphere radius may be obtained by inserting Eq. \eqref{new-metric} into Eq. \eqref{r0}. Furthermore, the radius of shadow circles for the higher-dimensional MOG dark copact object in celestial coordinates may be found by utilizing Eq. \eqref{Rs} and applying Eq. \eqref{new-metric}  into Eq. \eqref{xieta2}. In Table \ref{Table1} we collect the numerical data associated with $r_eh$, and $r_0$ for $D = 4, 5, 6, 7, 8$ and some different values of $\alpha$.
\begin{table}[htb]
        \centering
        \caption{Values of $r_{eh}$, and $r_{0}$ for different values of $\alpha$ and $D$.}
        \label{Table1}
        \begin{tabular}{|c||c|c||c|c||c|c||c|c||c|c|}
        \hline
             & \multicolumn{2}{|c||}{$D=4$} & \multicolumn{2}{|c||}{$D=5$} & \multicolumn{2}{|c||}{$D=6$} & \multicolumn{2}{|c||}{$D=7$} & \multicolumn{2}{|c|}{$D=8$}\\
             \cline{2-11}
          $D$   & $r_{eh}$ & $r_{0}$ & $r_{eh}$ & $r_{0}$ & $r_{eh}$ & $r_{0}$ & $r_{eh}$ & $r_{0}$ & $r_{eh}$ & $r_{0}$\\
            \hline\hline
            \multicolumn{1}{|c||}{$\alpha=0$} & 2 & 3 & 1.4142 & 2 & 1.25992 & 1.70998 & 1.1892 & 1.56508 & 1.14869 & 1.58482\\
            \hline
            \multicolumn{1}{|c||}{$\alpha=0.2$} & 2.29545 & 3.26804 & 1.51507 & 2.11584 & 1.31913 & 1.78537 & 1.23088 & 1.62095 & 1.18079 & 1.61927\\
            \hline
            \multicolumn{1}{|c||}{$\alpha=0.45$} & 2.65416 & 3.50244 & 1.62916 & 2.23837 & 1.38455 & 1.86748 & 1.27639 & 1.68194 & 1.21559 & 1.46514\\
            \hline
            \multicolumn{1}{|c||}{$\alpha=0.5$} & 2.72474 & 3.53269 & 1.65068 & 2.26026 & 1.39672 & 1.88259 & 1.28479 & 1.6932 & 1.22198 & 1.51352\\
            \hline
            \multicolumn{1}{|c||}{$\alpha=0.674$} & 2.96783 & 3.57489 & 1.72274 & 2.33008 & 1.43708 & 1.93236 & 1.31253 & 1.73039 & 1.24305 & 1.5618\\
            \hline
        \end{tabular}
\end{table}

\begin{figure}[htb]\label{fig3}
\centering
\subfloat[\label{Fig3a} for $\alpha=0.674$]{\includegraphics[width=0.32\textwidth]{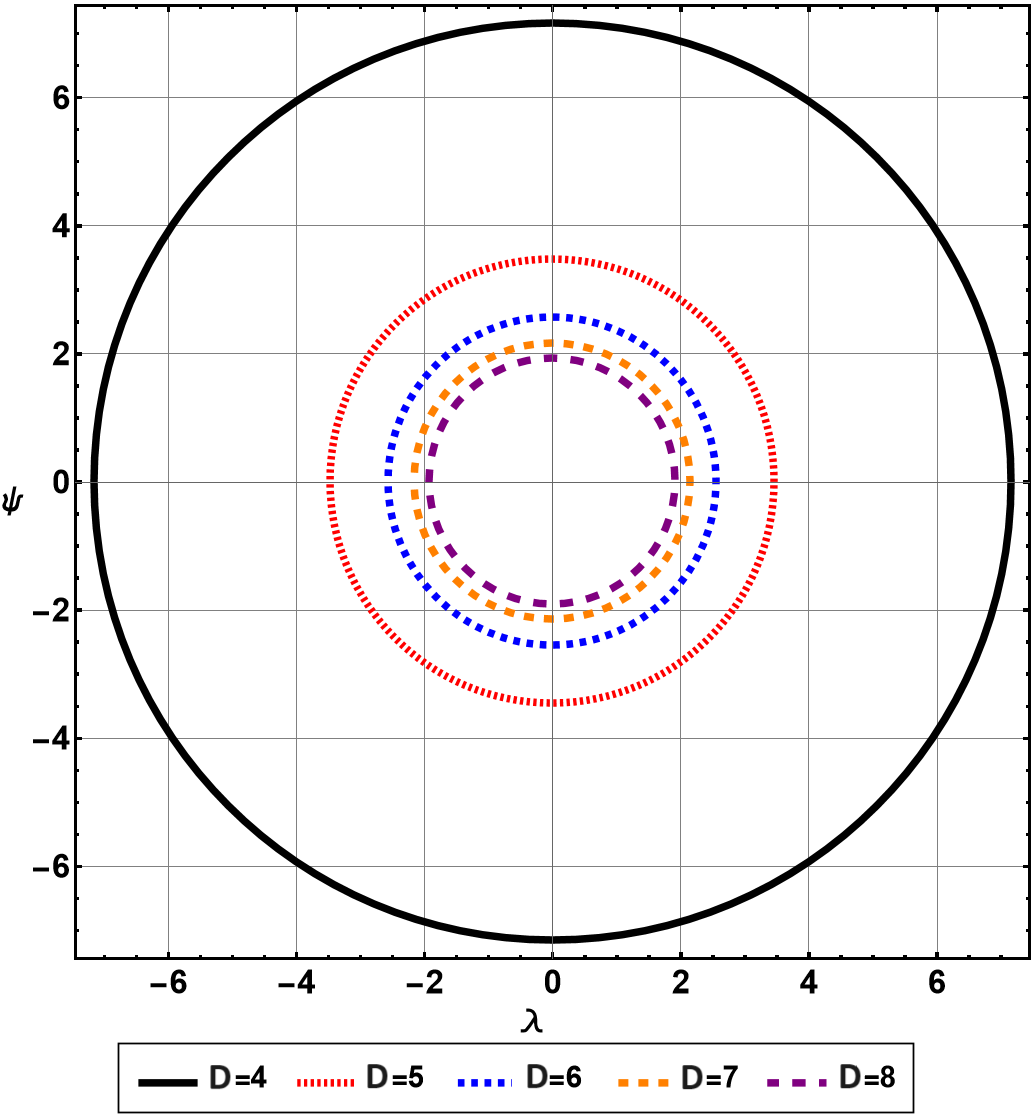}}
\,\,\,
\subfloat[\label{Fig3b} for $\alpha=0.5$]{\includegraphics[width=0.32\textwidth]{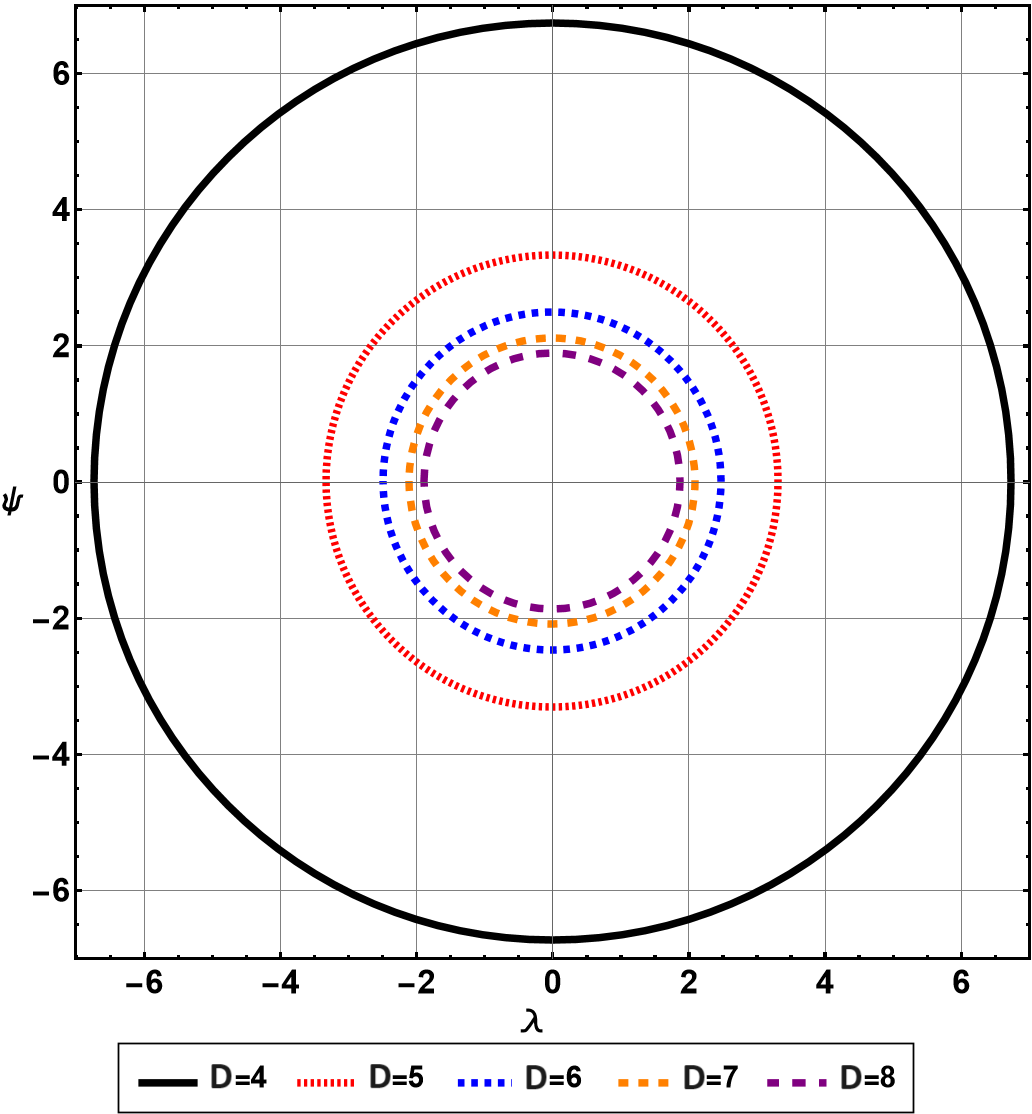}}
\,\,\,
\subfloat[\label{Fig3c} for $\alpha=0.45$]{\includegraphics[width=0.32\textwidth]{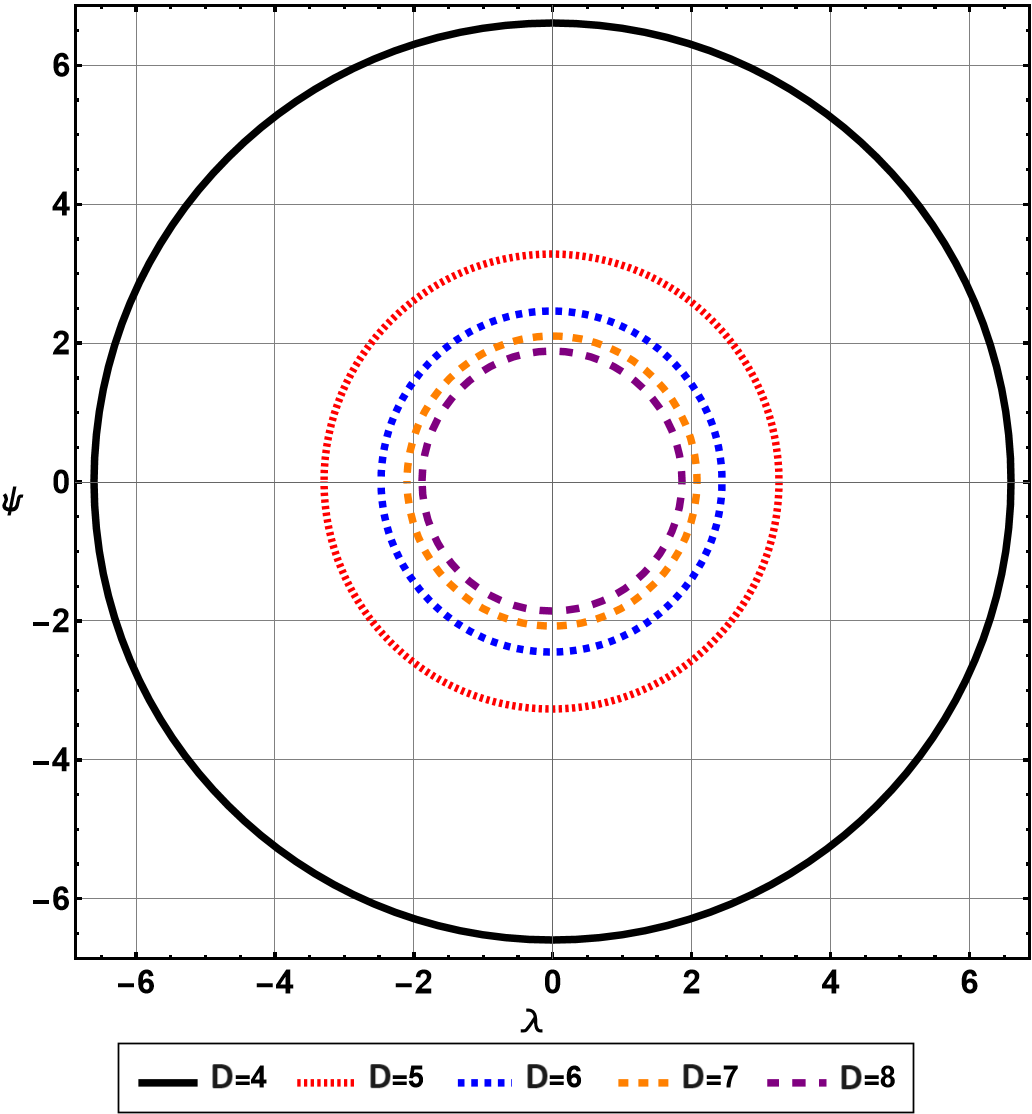}}
\\
\subfloat[\label{Fig3d} for $\alpha=0.2$]{\includegraphics[width=0.32\textwidth]{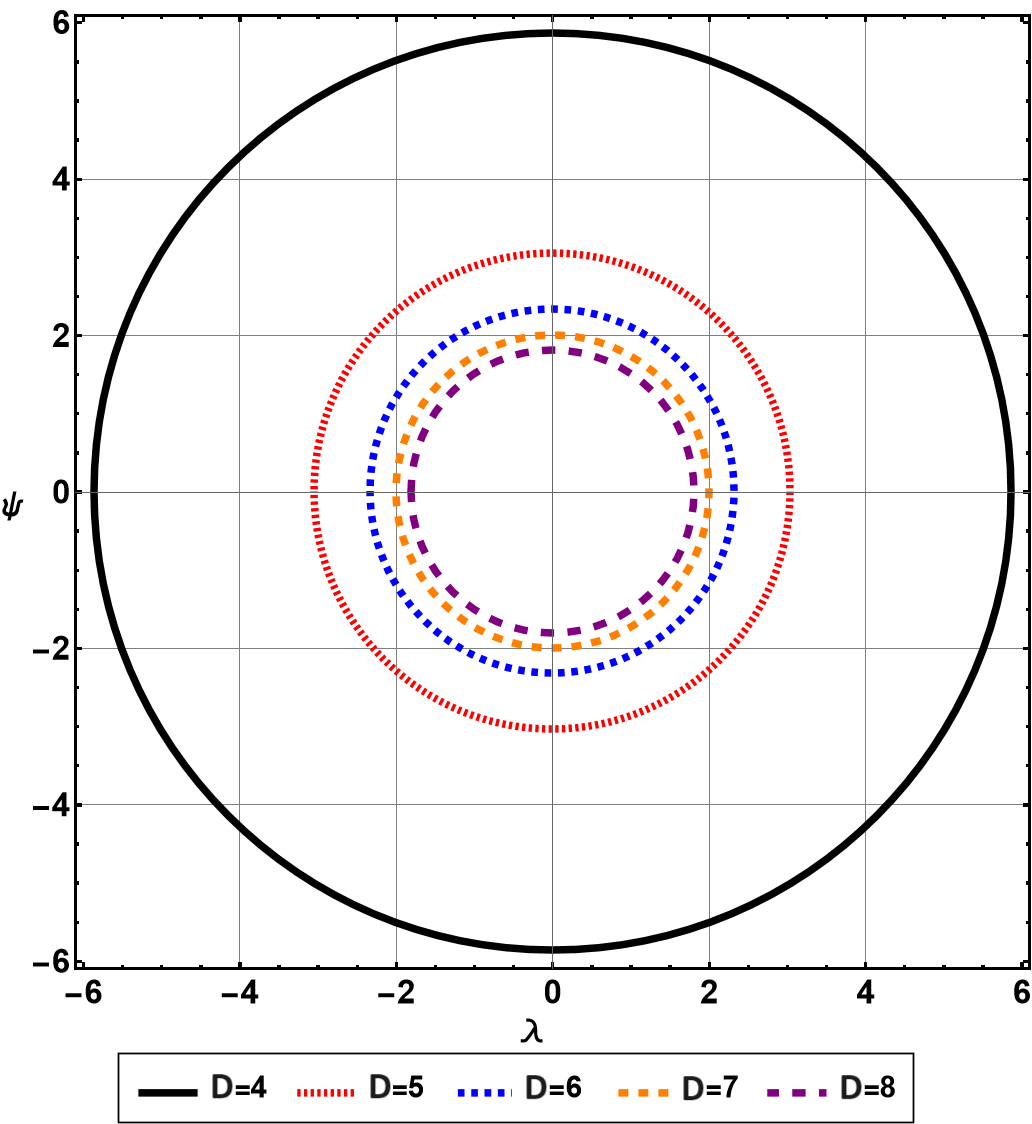}}
\,\,\,
\caption{\label{Fig3}\small{\emph{The geometrical shape of the shadow in celestial plane of the higher-dimensional MOG dark compact object with $M=1$.}}}
\end{figure}
Plotting the shadow circles of the MOG dark compact object with extra dimensions for various values of $\alpha$ and $D$ is possible using the information given in Table \ref{Table1}. Figure \ref{Fig3} shows the geometrical shapes of the MOG dark compact object's shadow with extra dimensions in celestial coordinates for various values of $\alpha$ and $D$. As one can see it shows plots for constant values of $\alpha$ and illustrates how the black hole's shadow shapes get smaller as dimension increases for a certain value of $\alpha$. Therefore, by reducing the size of shadow's geometrical shape, the extra dimensions have a major impact on the black hole's shadow.

\begin{figure}[htb]
\centering
\subfloat[\label{Fig10a} for $D=4$]{\includegraphics[width=0.32\textwidth]{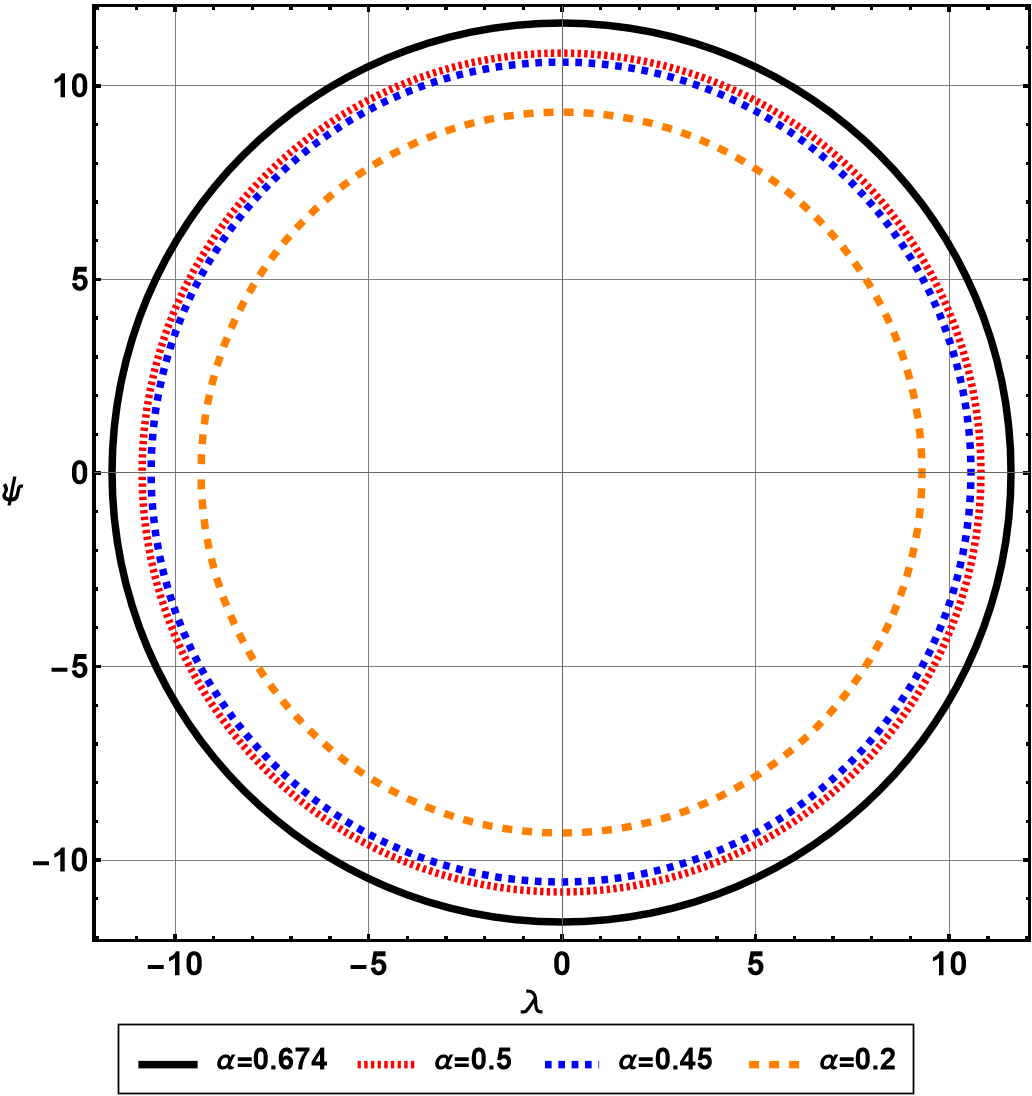}} \\
\subfloat[\label{Fig10b} for $D=5$]{\includegraphics[width=0.32\textwidth]{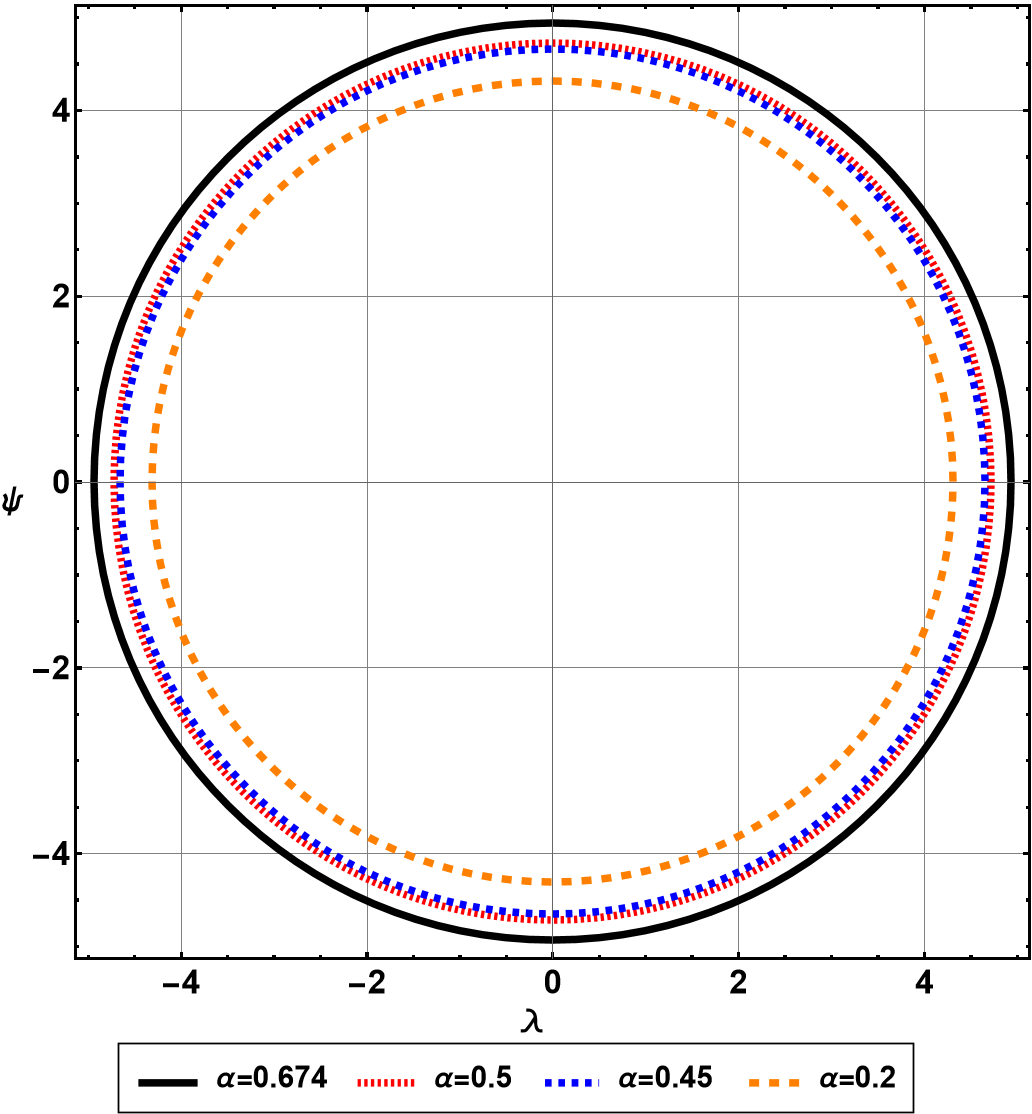}}
\,\,\,
\subfloat[\label{Fig10c} for $D=6$]{\includegraphics[width=0.32\textwidth]{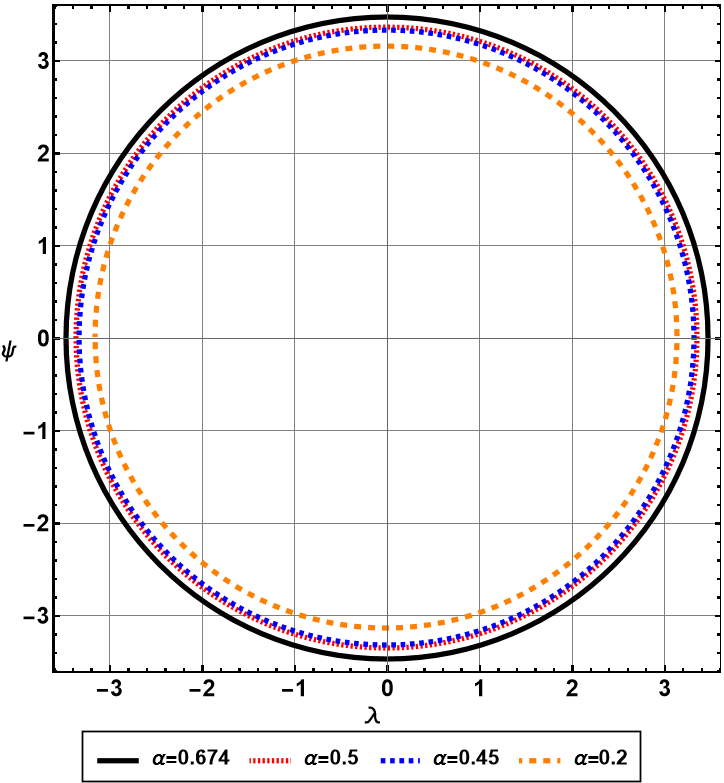}}
\caption{\label{Fig10}\small{\emph{The geometrical shape in celestial plane of the shadow of MOG dark compact object with higher dimensions with $M=1$.}}}
\end{figure}

Additionally, the MOG dark compact object's shadow circles with extra dimensions in the celestial coordinates for various values of $\alpha$ are visible in Fig.\ref{Fig10}. As one can see for $D=4$ in Fig. \ref{Fig10a} and $D=5,6$ in Figs \ref{Fig10b} and \ref{Fig10c}. As we can see in Fig. \ref{Fig10a} for $D=4$, the size of the shadow circles get larger as the STVG parameter value increases. However, in Figs. \ref{Fig10b} and \ref{Fig10c}, the MOG dark compact object's shadow size for $D=5,6$ with higher dimensions for each value of $\alpha$, approaching each other even as their sizes decrease relative to comparable ones for $D=4$. Additionally, the behavior for $D>6$ is obvious. Consequently, Fig. \ref{Fig10} demonstrates that the shadow circles corresponding to various values of $\alpha$ coincide on one another for $D\geq 6$. Consequently, with STVG theory, the effect of $\alpha$ on the higher-dimensional dark compact object's shadow is amplifier.

\subsection{Energy Emission rate}\label{sec31}

Hawking radiation is a phenomenon that allows black holes to radiate. The absorption cross-section typically oscillates around a limiting constant $\sigma_{lim}$ at very high energies. However, the absorption cross-section advances in the direction of the black hole shadow for an observer at a very far distance \cite{Wei:2013kza,Belhaj:2020rdb}. It is possible to demonstrate that $\sigma_{lim}$ is roughly equivalent to the area of the photon sphere, which may be expressed as follows in arbitrary dimension \cite{Wei:2013kza,Li:2020drn,Decanini:2011xw}
\begin{equation}\label{sigmalim}
\sigma_{lim}\approx\frac{\pi^{\frac{n-2}{2}}}{\Gamma\left[\frac{n}{2}\right]}R_{s}^{n-2}\,.
\end{equation}
 Form of the energy emission rate of higher-dimensional black holes is
\begin{equation}\label{radirate}
\frac{d^{2}E(\varpi)}{d\varpi dt}=\frac{2\pi^{2}\sigma_{lim}}{e^{\frac{\varpi}{T}}-1}\varpi^{D-1}\,,
\end{equation}
where $\varpi$ is the emission frequency and $T$ is the Hawking temperature of the black hole.

For getting the values of the Hawking temperature of the MOG dark compact object with extra dimensions for various values of $\alpha$ and $D$, we now use the values of $r_{eh}$ from Table \ref{Table1}. Next, by inserting the values of $\sqrt{\eta+\xi^{2}}$ into Eq. (\ref{sigmalim}), one can determine the values of $\sigma_{lim}$ for various values of $\alpha$ and $D$ which correspond to the higher-dimensional MOG dark compact object. In order to obtain the energy emission rate in terms of various values of $\alpha$ and $D$ related to the higher-dimensional MOG dark compact object, one can finally use the values of the Hawking temperature and $\sigma_{lim}$ in Eq. \eqref{radirate}.

\begin{figure}[htb]
\centering
\subfloat[\label{Fig4a} for $\alpha=0.45$]{\includegraphics[width=0.48\textwidth]{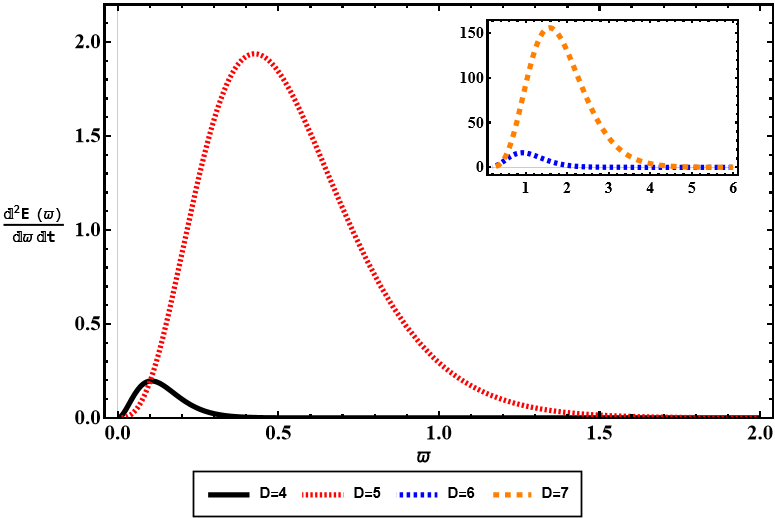}}
\,\,\,
\subfloat[\label{Fig4b} for $\alpha=0.2$]{\includegraphics[width=0.499\textwidth]{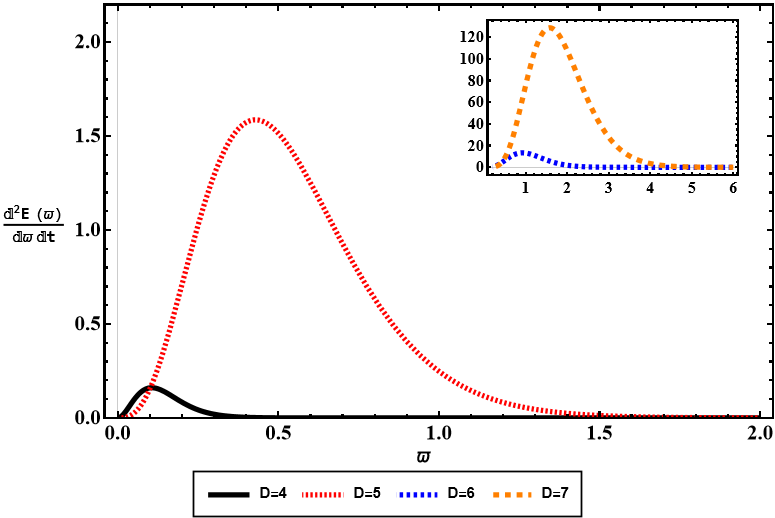}} \\
\subfloat[\label{Fig4c} for  $D=4$]{\includegraphics[width=0.49\textwidth]{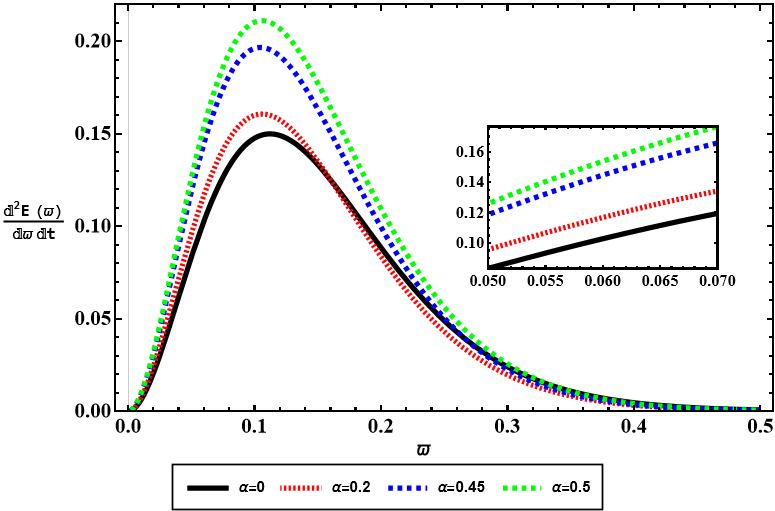}}
\,\,\,
\subfloat[\label{Fig4d} for $D=5$]{\includegraphics[width=0.49\textwidth]{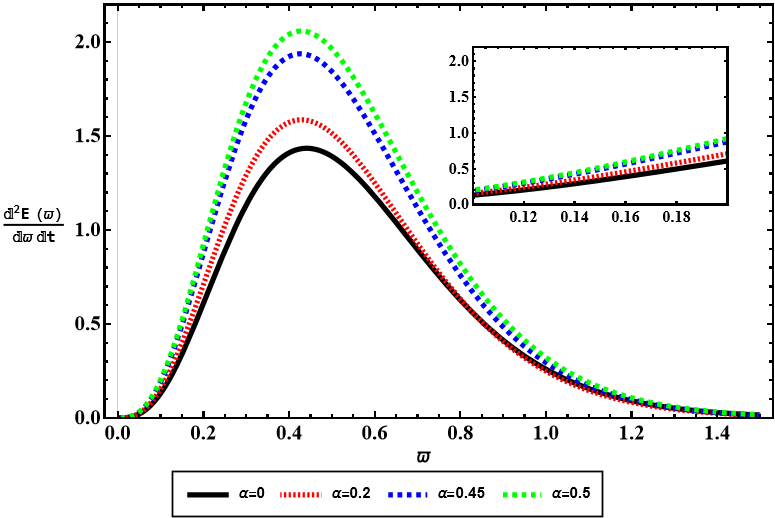}}
\caption{\label{Fig4}\small{\emph{The energy emission rate as a function of $\varpi$ for different values of $D$ and $\alpha$.}}}
\end{figure}
Fig. \ref{Fig4} shows the energy emission rate as a function of the emission frequency for the higher-dimensional MOG dark compact object. The energy emission rate of MOG dark compact object with extra dimensions greatly increases by growing $D$, as shown in Figs. \ref{Fig4a} and \ref{Fig4b}, at a fixed value of $\alpha$. Consequently, we discover that in STVG with higher dimensions, extra dimensions speed up the MOG compact object's evaporation. Also, the STVG parameter would increase the energy emission rate of MOG dark compact object, which causes
it's evaporation to accelerate.

\subsection{Deflection Angle}\label{sec32}

In this section a setting for examining the deflection angle of higher-dimensional black holes in the line element \eqref{le} is given. We intend to apply the Gauss-Bonnet theorem \cite{Gibbons:2008rj,Arakida:2017hrm}. To begin, we compute the optical metric on the equatorial hyperplane $\theta_{i}=\pi/2$ using the line element \eqref{le}. On this hyperplane, we set $d\theta_{D-2}^{2}=d\phi^{2}$ to find
\begin{equation}\label{optmet1}
ds^{2}=-h(r)dt^{2}+\frac{dr^{2}}{f(r)}+r^{2}d\phi^{2}\,.
\end{equation}
Then, for the considered null geodesics for which $ds^{2}=0$, the optical metric reads as follows
\begin{equation}\label{optmet2}
dt^{2}=\frac{dr^{2}}{h(r)f(r)}+\frac{r^{2}}{h(r)}d\phi^{2}\,.
\end{equation}
For this optical metric, we can calculate the Gaussian optical curvature $K=\frac{\bar{R}}{2}$ in which $\bar{R}$ is the Ricci scalar of the metric \eqref{optmet2} as follows
\begin{equation}\label{gauoptcur1}
K=\frac{2\,rh(r)f(r)h''(r)-2\,rf(r)h'^{2}(r)+h(r)h'(r)\left\{rf'(r)+2f(r)\right\}-2f'(r)h^{2}(r)}{2\,rh(r)}\,.
\end{equation}

In order to calculate the deflection angle, one should consider a non-singular manifold $\mathcal{D}_{\tilde{R}}$ with a geometrical size $\tilde{R}$ to employ the Gauss-Bonnet theorem, so that \cite{Gibbons:2008rj,Arakida:2017hrm}
\begin{equation}\label{gaubon1}
\int\int_{\mathcal{D}_{\tilde{R}}}KdS+\oint_{\partial\mathcal{D}_{\tilde{R}}}kdt+\sum_{i}\varphi_{i}=2\pi\zeta(\mathcal{D}_{\tilde{R}})\,,
\end{equation}
where $dS=\sqrt{\bar{g}}drd\phi$ and $dt$ are the surface and line element of the optical metric \eqref{optmet2}, respectively, $\bar{g}$ is the determinant of the optical metric, $k$ denotes the geodesic curvature of $\mathcal{D}_{\tilde{R}}$, and $\varphi_{i}$ is the jump (exterior) angle at the $i$-th vertex, and also, $\zeta(\mathcal{D}_{\tilde{R}})$ is the Euler characteristic number of $\mathcal{D}_{\tilde{R}}$. One can set $\zeta(\mathcal{D}_{\tilde{R}})=1$. Then, considering a smooth curve $y$, which has the tangent vector $\dot{y}$ and acceleration vector $\ddot{y}$, the geodesic curvature $k$ of $y$ can be defined as follows where the unit speed condition $\tilde{g}\left(\dot{y},\dot{y}\right)=1$ is employed
\begin{equation}\label{geocur}
k=\tilde{g}\left(\nabla_{\dot{y}\dot{y},\ddot{y}}\right)\,,
\end{equation}
which is a measure of deviations of $y$ from being a geodesic. In the limit $\tilde{R}\rightarrow\infty$, two jump angles $\varphi_{s}$ (of source) and $\varphi_{o}$ (of observer) will become $\pi/2$, i.e, $\varphi_{s}+\varphi_{o}\rightarrow\pi$. Considering $C_{\tilde{R}}:=r(\phi)$, we have $k(C_{\tilde{R}})=|\nabla_{\dot{C}_{\tilde{R}}}\dot{C}_{\tilde{R}}|\,\,\,\myeqq\,\,\, 1/\tilde{R}$ and therefore, we can find $\lim_{\tilde{R}\rightarrow\infty}dt=\tilde{R}d\phi$. Hence, $k(C_{\tilde{R}})dt=d\phi$. Consequently, the Gauss-Bonnet theorem will reduce to the following form
\begin{equation}\label{gaubon2}
\int\int_{\mathcal{D}_{\tilde{R}}}KdS+\oint_{C_{\tilde{R}}}kdt\,\,\,\myeq\,\,\,\int\int_{\mathcal{D}_{\infty}}KdS+\int_{0}^{\pi+\Theta}d\phi=\pi\,.
\end{equation}
Thus, following equation for calculating the deflection angle (see Refs. \cite{Gibbons:2008rj,Arakida:2017hrm} and references therein)
\begin{equation}\label{defang}
\Theta=\pi-\int_{0}^{\pi+\Theta}d\phi=-\int_{0}^{\pi}\int_{\frac{\xi}{\sin\phi}}^{\infty}KdS\,.
\end{equation}

Applying Eq. (\ref{new-metric}) into Eqs. (\ref{optmet2}) and (\ref{gauoptcur1}), yields the Gaussian optical curvature for the higher dimensional MOG dark compact object
\begin{equation}\label{kstvg}
K\approx \frac{21M(D-3)}{2}\bigg\{\Big[\frac{2((D-2)\alpha+\frac{D}{3}-\frac{1}{3})(1+\alpha)Mr^{(5-2D)}}{7}\Big]-\Big[\frac{2(1+\alpha)(D-2)r^{(2-D)}}{21}\Big]+\Big[\alpha M^{2}r^{(8-3D)}(D-\frac{20}{7})\Big]\bigg\}
\end{equation}
Furthermore, the surface element of the optical metric (\ref{optmet1}) for the higher-dimensional MOG dark compact object in
correspondance with the metric coefficients (\ref{new-metric}) is approximately 

\begin{equation}\label{surele1}
dS=\sqrt{\bar{g}}\,drd\phi=\frac{r}{h(r)\sqrt{f(r)}}dtd\phi\approx rdrd\phi\,.
\end{equation}

 The deflection angle of the higher-dimensional MOG dark compact object as follows

\begin{equation}\label{defangstvg}
\begin{split}
\Theta & =-\int_{0}^{\pi}\int_{\frac{\xi}{\sin\phi}}^{\infty}KdS\\
& \approx-\int_{0}^{\pi}\int_{\frac{\xi}{\sin\phi}}^{\infty}\frac{21M(D-3)}{2}\bigg\{\Big[\frac{2((D-2)\alpha+\frac{D}{3}-\frac{1}{3})(1+\alpha)Mr^{(5-2D)}}{7}\Big]\\
&
-\Big[\frac{2(1+\alpha)(D-2)r^{(2-D)}}{21}\Big]+\Big[\alpha M^{2}r^{(8-3D)}(D-\frac{20}{7})\Big]\bigg\}
=\frac{M \sqrt{\pi}}{\xi^{D-4}}\bigg\{\frac{4(D-2)(1+\alpha)\Gamma[\frac{D-1}{2}]}{(D-4)^{2}\Gamma[\frac{D-4}{2}]}+\frac{(20-7D)M^{2}\alpha \Gamma[\frac{3D-7}{2}]}{\xi^{(2D-6)}(3D-10)\Gamma[\frac{3D-8}{2}]}\bigg\}\,.
\end{split}
\end{equation}

\begin{figure}[htb]
\centering
\subfloat[\label{defa} for $\alpha=0.45$]{\includegraphics[width=0.49\textwidth]{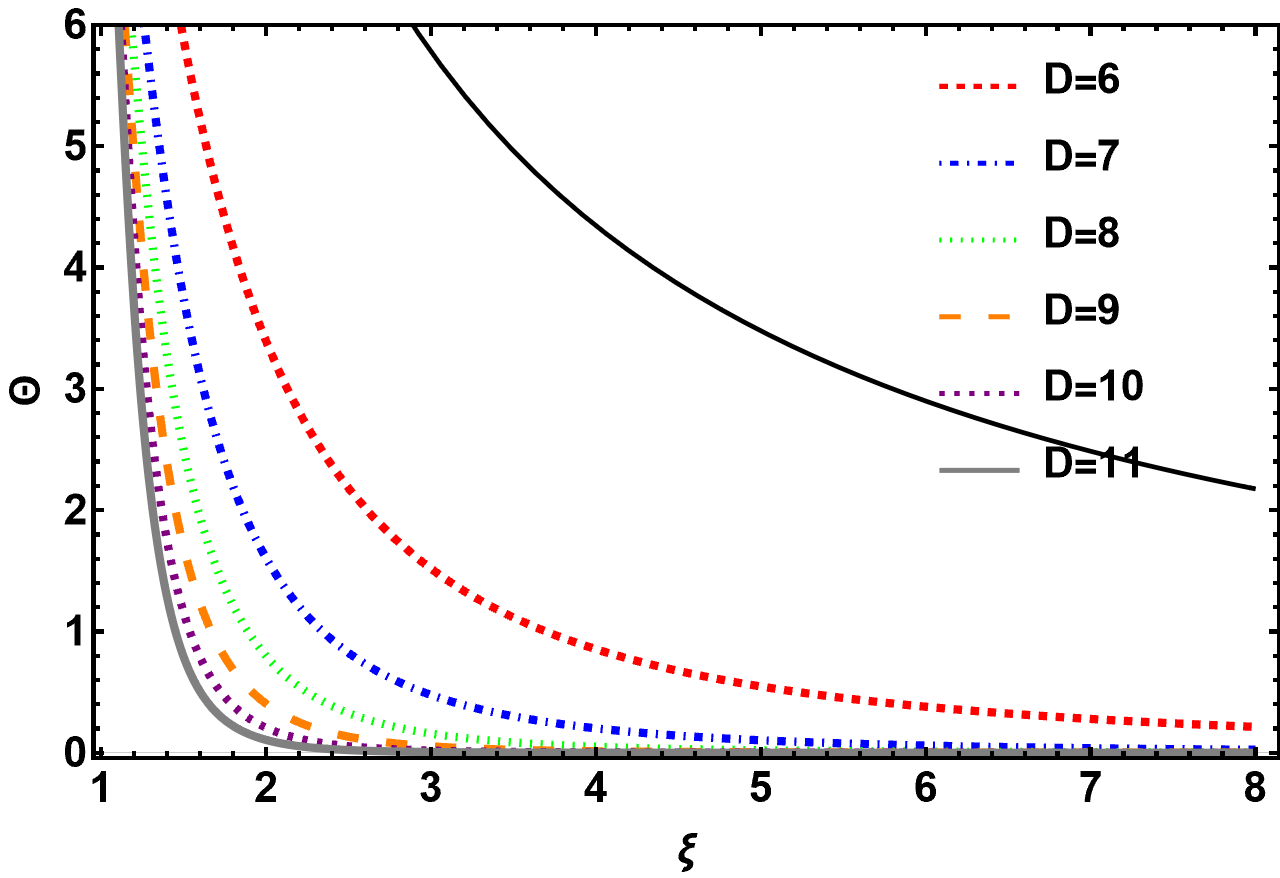}} \\
\subfloat[\label{defb} for $D=5$]{\includegraphics[width=0.49\textwidth]{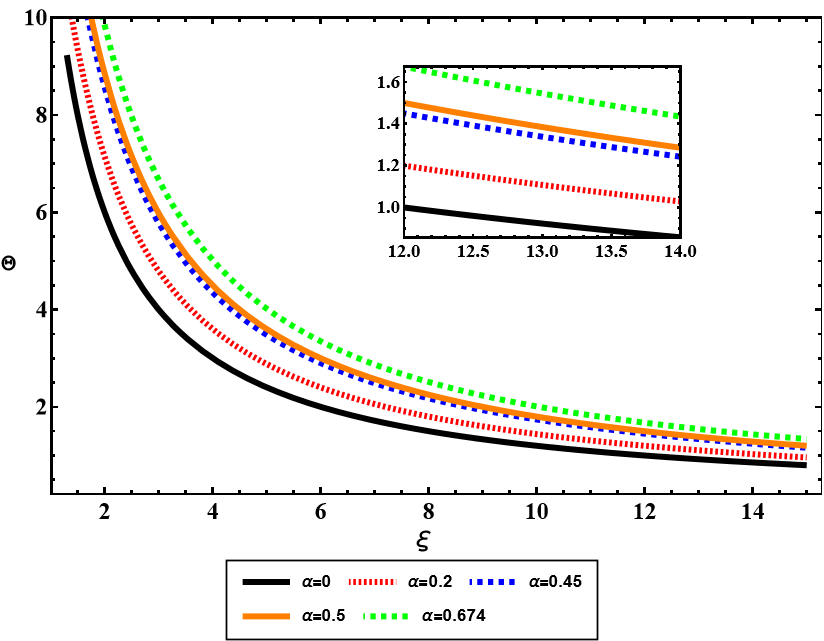}}
\,\,\,
\subfloat[\label{defc} for $D=6$]{\includegraphics[width=0.49\textwidth]{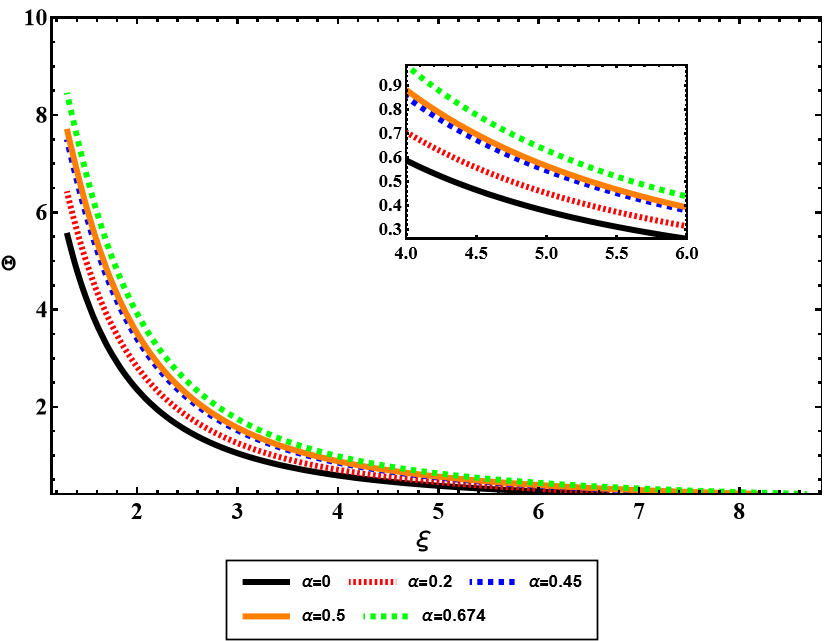}}
\caption{\label{Fig6}\small{\emph{The behavior of the deflection angle of the higher-dimensional MOG dark compact object in terms of $\xi$ for different values of $D$ and $\alpha$..}}}
\end{figure}

The behavior of the deflection angle of the higher-dimensional MOG dark compact object is illustrated in Fig. \ref{Fig6} for different values of $D$ with respect to $\alpha=0.45$ in Fig. \ref{defa} and for different values of $\alpha$ with respect to $D=4, 5$ in Figs. \ref{defb} and \ref{defc}, respectively. Fig. 7a, shows us that decreasing the value of the impact parameter $\xi$ results in extremely
increasing the deflection angle of the MOG dark compact object. From Fig. \ref{defa}, it is obvious that for a constant value of STVG parameter, the deflection angle of the MOG dark compact object reduces by growing the number of extra dimensions. In Fig. \ref{defb} and Fig. \ref{defc}  it is clear that for a fixed value of $D$,increasing STVG parameter $\alpha$ would result in decreasing the deflection angle.

\section{Constraining STVG Parameter from EHT observations of M87* }\label{sec4}

Here we aim to compare the deduced shadow radius of the higher-dimensional MOG dark compact object with the shadow size of supermassive black hole, M87* captured by EHT. Within $1$-$\sigma$ (68\%) confidence level, one can find that the shadow size of M87* supermassive black hole captured by EHT lies within the interval \cite{EventHorizonTelescope:2021dqv}
\begin{equation}\label{M87}
4.31\leq R_{s, M87^{*}}\leq 6.08\,.
\end{equation}
Comparing this with the shadow size of the higher-dimensional MOG dark compact object enables us to constrain the STVG parameter $\alpha$.

\begin{figure}[htb]
\centering
\includegraphics[width=0.7\textwidth]{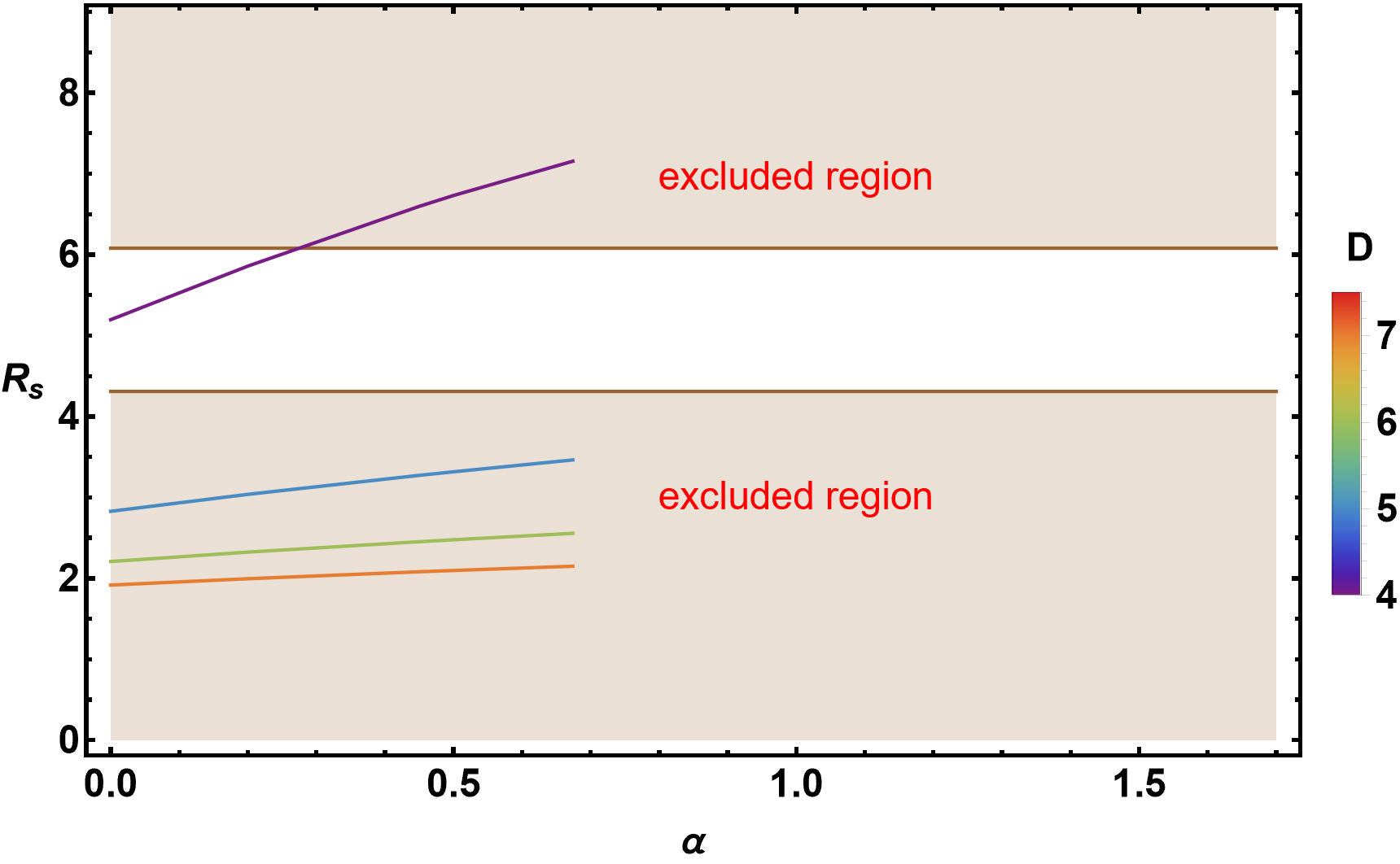}
\caption{\small{\emph{The shadow radius of the extra-dimensional MOG dark compact object, compared to the shadow size of $M87^{*}$ captured by EHT within $1-\sigma$ confidence level versus the parameter $\alpha$. The colored area is excluded region, which is inconsistent with the EHT data, while the white region is the $1-\sigma$ confidence level of EHT data.}}}
\label{fig7}
\end{figure}

Figure \ref{fig7} indicates the behavior of the shadow radius of the higher-dimensional MOG dark compact object in comparison to the EHT's shadow size of M87* within $1$-$\sigma$ uncertainties as reported in Eq. \eqref{M87} versus the STVG parameter $\alpha$. In this figure, it is so obvious that the shadows of the higher-dimensional MOG dark compact object associated with $D=5,6,7$ are incompatible with the observations of EHT. However, the shadow of the MOG dark compact object with $D=4$ is inside the $1$-$\sigma$ confidence level of EHT data, and in the limit $0\leq \alpha<0.279$ the shadow radius of the four dimensional MOG dark compact object demonstrates good concordance with EHT observations. Additionally, such as in Table \ref{Table1} and Fig. \ref{Fig3a}, it is clear from Fig. \ref{fig7} that increasing the STVG parameter value results in amplifying the shadow radius of the four dimensional MOG dark compact object. For $D\geq5$ the effect of the STVG parameter in comparison to the extra dimension impact is somehow negligible and the effect of extra dimensions is dominated as it is obvious in Fig. \ref{fig7}

\section{Acceleration bounds for higher-dimensional MOG dark compact object}\label{sec5}

As an important issue in this setup, in this section we study the correlation between the shadow radius of the higher-dimensional MOG dark compact object and the acceleration bounds on radial linear uniformly accelerated (LUA) trajectories, i.e. the
curved spacetime equivalent of the hyperbolic Rindler trajectories in a flat spacetime. The Rindler trajectories in
flat spacetime can be described in a covariant way using the Letaw-Frenet equations, which involve three geometric
scalars for a curve. For these trajectories, only the curvature scalar is non-zero and constant, and it is equal to the
acceleration magnitude. The other two scalars, torsion and hyper-torsion, are zero. This covariant definition
can be extended to curved spacetime, and the resulting curves with constant curvature scalar and zero torsion and
hyper-torsion scalars are locally hyperbolic and linear in a local inertial frame at any point on the curve.
From the linearity condition for LUA, the uniform acceleration can be expressed as the following equation

\begin{equation} \label{acc}
u^{j}\nabla_{j}a^{i}-a^{2}u^{i}=0,
\end{equation}
where $a^{i}$ and $u^{i}$ are respectively the acceleration and velocity vectors along the trajectory. By solving the above
equations for a given spacetime, one can find the set of LUA trajectories with the constant magnitude of acceleration $|a|$. Therefore, we obtain LUA trajectory consistent with the constraint of the equations (\ref{acc}) by considering the higher-dimensional MOG dark compact object metric (\ref{new-metric})

\begin{equation}
\frac{dt}{d\lambda}=f^{-1}(r)(|a|r+h),
\end{equation}

\begin{equation}
\frac{dr}{d\lambda}=\pm \sqrt{(|a|r+h)^{2}-f(r)},
\end{equation}
where $\lambda$ is the proper time along the trajectory, $|a|$ is the uniform acceleration magnitude and $h$ is defined as a
boundary data. In the case where $h = 0$, the radial LUA trajectories follow the relation

\begin{equation}
a^{2}r^{2}-1-\frac{M^{2}r^{2(D-3)}\alpha(1+\alpha)}{(r^{2(D-3)}+M^{2}\alpha (1+\alpha))^{2}} + \frac{2Mr^{2(D-3)}(1+\alpha)}{(r^{2(D-3)}+M^{2}\alpha (1+ \alpha))^{\frac{3}{2}}}=-V_{eff}
\end{equation}
in which $|a|\neq0$. The above equation specifies the kinematics of the radial LUA trajectory. As a condition at turning
point $r_{t}$ of this trajectory, the radial velocity should be zero which constrains the effective potential to be zero. Hence,

\begin{equation}\label{acc1}
-V_{eff}=a^2 r_{t}^2 \left(r_{t}^{2 D-6}+\alpha  (\alpha +1) M^2\right)^{3/2}-\left(r_{t}^{2 D-6}+\alpha  (\alpha +1) M^2\right)^{3/2}-\frac{\alpha  (\alpha +1) M^2 r_{t}^{2 D-6}}{\sqrt{a^{2 D-6}+\alpha  (\alpha +1) M^2}}+2 (\alpha +1) M r_{t}^{2 D-6}=0
\end{equation}

At the bound $|a| = |a|_b$, the turning point is an extremum of the above equation, thus we have

\begin{equation}\label{acc2}
\begin{split}
0=\frac{1}{r_b^7} & \bigg\{r_b^{2 D}\Big
[(3 D-7) a_b^2 r_b^2-3 D+9\Big] \sqrt{r_b^{2 D-6}+\alpha  (\alpha +1) M^2}+4 (\alpha +1) (D-3) M r_b^{2 D}+\frac{1}{\left(r_b^{2 D}+\alpha  (\alpha +1) M^2 r_b^6\right)^2}\\
& \times\bigg(\alpha  (\alpha +1) M^2 r_b^6 \sqrt{r_b^{2 D-6}+\alpha  (\alpha +1) M^2} \Big[2 \alpha  (\alpha +1) M^2 r_b^{2 D+6} \left(2 a_b^2 r_b^2-D+3\right)+r_b^{4 D} \left(2 a_b^2 r_b^2-D+3\right)\\
& +2 \alpha ^2 (\alpha +1)^2 M^4 a_b^2 r_b^{14}\Big]\bigg)\bigg\}
\end{split}
\end{equation}

By solving equations (\ref{acc2}) and (\ref{acc1}) simultaneously, one can find the following equations for the turning point

\begin{equation}\label{acc3}
|a|_{b}=\frac{\sqrt{\frac{r_{b}^{4D}+M^{4}r_{b}^{12}\alpha ^{2}(1+\alpha)^{2}+Mr_{b}^{(6+2D)}(1+\alpha)(3M\alpha -2\sqrt{r_{b}^{(2D-6)}+M^{2}\alpha(1+\alpha)})}{r_{b}^{12}\sqrt{r_{b}^{(2D-6)}+M^{2}\alpha(1+\alpha)}}}}{r_{b}(r_{b}^{(2D-6)}+M^{2}\alpha(1+\alpha))^{\frac{3}{4}}}
\end{equation}

\begin{equation}\label{acc4}
\begin{split}
0=-\frac{2}{\left(r^{2 D}+\alpha  (\alpha +1) M^2 r^6\right)^3} & \bigg\{(\alpha +1) M r^{4 D+6} \left(\alpha  (D+1) M-(D-1) \sqrt{r^{2 D-6}+\alpha  (\alpha +1) M^2}\right)\\
& +\alpha  (\alpha +1)^2 M^3 r^{2 (D+6)} \left(2 (D-4) \sqrt{r^{2 D-6}+\alpha  (\alpha +1) M^2}-\alpha  (D-7) M\right)\\
& +r^{6 D}+\alpha ^3 (\alpha +1)^3 M^6 r^{18}\bigg\}
\end{split}
\end{equation}

Since the equations (\ref{acc4}) and (\ref{potential2}) are the same, the bound on the distance of closest approach $r_b$ will be equal to the
radius of photon sphere $r_0$. By setting $r_b = r_0$ and doing some calculations, one can find out that the equation (\ref{acc3})
is related to the shadow radius of the higher-dimensional MOG dark compact object as follows

\begin{equation}
|a|_{b}=\frac{1}{R_s}
\end{equation}

In what follows we provide numerical results of the acceleration bound for LUA trajectory in a higher-dimensional MOG dark compact object background. As we expected, the acceleration bound $|a|_b$ is the reverse of the value of the shadow radius given
in subsection \ref{GEO}.

\begin{table}[htb]
\centering
\caption{\label{Table3}\small{\emph{The numerical values of acceleration bound $|a|_{b}$ for LUA trajectory in a higher-dimensional MOG dark compact object
spacetime for some different values of the parameters $\alpha$ and $D$.}}}\renewcommand{\arraystretch}{1.3}
\begin{tabular}{|c|c|c|c|}
\hline
\multicolumn{1}{|c|}{\multirow{2}{*}{$\alpha$}} & \multicolumn{3}{c|}{$|a|_{b}$}        \\ \cline{2-4}
\multicolumn{1}{|c|}{}                          & $D=4$      & $D=7$      & $D=8$      \\ \hline
$0.674$                                       & $0.139761$ & $0.379497$ & $0.449613$ \\ \hline
$0.45$                                        & $0.151589$ & $0.395799$ & $0.446273$ \\ \hline
$0.2$                                         & $0.170696$ & $0.418589$ & $0.50117$  \\ \hline
\end{tabular}
\end{table}

\section{Concluding Remarks}\label{sec6}

In this study, according to string theory, braneworld models, and AdS/CFT correspondence, we motivated to obtain a higher-dimensional MOG dark compact object solution. We investigated the behaviors of the corresponding MOG dark compact object's shadow and deflection angle. Determining how extra dimensions and the STVG parameter affect the shadow of the black holes was our main concern. At first we developed a higher-dimensional metric for a dark compact object in STVG theory, then by using the Hamilton-Jacobi approach and Carter method to develop the null geodesics around them and calculate the corresponding effective potentials. After that, we used the celestial coordinates to determine the shadow shape of the higher-dimensional STVG black holes on the observer's sky. In this higher-dimensional STVG framework, we also developed the formulas for the energy emission rate and deflection angle . The shadow behavior, deflection angle, and energy emission rate of a MOG dark compact object with extra dimensions were then investigated using the framework we developed. We computed and investigated the major effects of the STVG parameter and additional dimensions on the shadow, deflection angle, and energy emission rate of the black holes in this scenario. Furthermore, we restrict these parameters by comparing the shadow size of M87* obtained from EHT observations to the shadow radius of the higher-dimensional MOG dark compact object. Finally, the issue of acceleration bounds in higher-dimensional MOG dark compact object has been studied in this context in confrontation with EHT data of M87*.

For the higher-dimensional MOG dark compact object, we found that for a constant value of the STVG parameter $\alpha$, the shadow size of the MOG dark compact object decreases with increasing the number of extra dimensions $D$. Also, when the STVG parameter value increases with a fixed $D$, the shadow size of the MOG dark compact object increases. Also, we saw that for a fixed value of $\alpha$, the energy emission rate of the MOG dark compact object with higher dimensions extremely goes up by increasing $D$. Furthermore, increasing the STVG parameter results in growing the energy emission rate, not as mush as the extra dimensions. Therefore, we found that extra dimensions accelerate the evaporation of the MOG dark compact object with extra dimensions. After that, utilizing the Gauss-Bonnet theorem, we determined the deflection angle's leading terms in the weak-limit approximation. We've explored how the STVG parameter and extra dimensions affect this optical quantity. We noticed that for a constant value of the STVG parameter, the deflection angle of the MOG dark compact object with higher dimensions would reduce after we grow the number of dimensions. Also, for a fixed value of $D$, the deflection angle of the MOG dark compact object decreases by increasing the STVG parameter $\alpha$. Furthermore, we found that only the shadow of a four-dimensional MOG dark compact object with $0\leq \alpha<0.279$ is inside the $1$-$\sigma$ uncertainty of EHT data.

In summary, we can came to conclusion that the shadow of higher-dimensional MOG dark compact object is defined by the extra dimensions and also the STVG parameter. In this sense, the extra dimensions in STVG theory have a substantial effect on the shadows of the dark compact object, lowering their size. On the one hand, it was shown that it seems reasonable to detect the effects of the STVG theory via EHT. The crucial case here is that Fig. \ref{fig7} shows that the limit $\alpha<0.279$ can be noticed from the shadows of black holes caught by EHT. These findings may be resulted in the potential of testing the modified theories of gravity, and their solutions and their parameters by utilizing astrophysical data.

\begin{acknowledgments}

The authors would like to thank John W. Moffat for their fruitful comments and discussions, which improved the
quality of the paper, considerably.

\end{acknowledgments}

\end{document}